\documentclass[useAMS,usenatbib]{mn2e}
\usepackage{aas_macros}
\usepackage{natbib}
\usepackage{amssymb}
\usepackage{threeparttable}  
\usepackage{booktabs}
\usepackage{multirow}
\usepackage{bm}
\usepackage{amsfonts}
\usepackage{graphicx}
\usepackage{epstopdf}
\usepackage{longtable}
\usepackage{setspace}
\bibpunct[, ]{(}{)}{,}{a}{}{,}

\newcommand{\Eexc}{$E_{\rm exc}$}
\newcommand{\Teff}{$T_{\rm eff}$}  
\newcommand{\kms}{km\,s$^{-1}$}

\title[Monthly Notices ]
  {Carbon abundances of the reference late-type stars from 1D analysis of atomic C~I and molecular CH lines}
\author[S.~A.~Alexeeva $\&$  L.~I.~Mashonkina]
  {S.~A.~Alexeeva,$^1$ $\&$  L.~I.~Mashonkina,$^1$ 
  \newauthor 
  $^1$Institute of Astronomy of the Russian Academy of Sciences, \\
    48 Pyatnitskaya St. 119017, Moscow, Russia}
\date{Released 2014 Xxxxx XX}

\pagerange{\pageref{firstpage}--\pageref{lastpage}} \pubyear{2014}

\def\LaTeX{L\kern-.36em\raise.3ex\hbox{a}\kern-.15em
    T\kern-.1667em\lower.7ex\hbox{E}\kern-.125emX}

\begin{document}

\label{firstpage}

\maketitle

\begin{abstract}

A comprehensive model atom was constructed for C~I using the most up-to-date atomic data.
 We evaluated non-local thermodynamical equilibrium (NLTE) line formation for neutral carbon 
in classical 1D models representing atmospheres of late-type stars, where carbon abundance varies from solar value down to [C/H] = $-3$. NLTE leads to stronger C~I lines compared with their LTE strength and negative NLTE abundance corrections, $\Delta_{\rm NLTE}$. 
The deviations from LTE are large for the strong lines in the infrared (IR), with $\Delta_{\rm NLTE}$ = $-0.10$~dex to $-0.45$~dex
depending on stellar parameters, and they are minor for the weak lines in the visible spectral range, with $|\Delta_{\rm NLTE}| \le$ 0.03~dex. 
The NLTE abundance corrections were found to be dependent of the carbon abundance in the model. 
As the first application of the treated model atom, carbon NLTE abundances were determined for the Sun and eight late-type stars with well-determined stellar parameters that cover the $-2.56 \le$ [Fe/H] $\le -1.02$ metallicity range. Consistent abundances from the visible and IR lines were found for the Sun and the most metal-rich star of our sample, when applying a scaling factor of $S_{\rm H}$ = 0.3 to the Drawinian rates of C+H collisions.
Carbon abundances were also derived from the molecular CH lines and, for each star, they agree with that from the atomic C~I lines.
We present the NLTE abundance corrections for lines of C~I in the grid of model atmospheres applicable to the carbon-enhanced (CEMP) stars.

\end{abstract}


\section{Introduction}

Carbon abundance studies are of high importance in various fields of astrophysics and astrochemistry.
Carbon is an essential element for the beginnings of life on the Earth.
It acts as a primary catalyst for H-burning via the CNO cycle, and it contributes significantly to the stellar interior and atmosphere opacity.
Carbon plays an important role in dust formation processes in the interstellar medium.

Carbon is one of the representatives of the 'heavy' ($Z \ge 6$) elements, which are of stellar nucleosynthesis origin. Accurate stellar carbon abundances may improve our understanding the chemical evolution history of our Galaxy and other galaxies.
In the course of normal stellar evolution, carbon is essentially all produced by He burning through the 3$\alpha$ nuclear reaction: 3$^4$He $\to$ $^{12}$C $+$ $\gamma$ \citep{1957RvMP...29..547B}.  
In the modern galacto-chemical evolution models
the major sources of carbon are supernovae of type II (SNeII), hypernovae (HNe), supernovae of type Ia (SNeIa), and asymptotic giant branch (AGB) stars \citep[see, for example][]{2011MNRAS.414.3231K}. 
A relative contribution of different sources to the galactic carbon varied with time.
 
In the early Galaxy, the carbon production was dominated by SNeII and HNe. Indeed, progenitors of SNeII and HNe are massive stars, with a mass of larger than 8 and 20 solar mass (M$_\odot$), respectively, and a short lifetime of few 10$^6$--10$^7$ years. One also considers the scenarios with rotating massive stars. The delay time for nucleosynthesis in SNeIa ranges between 0.3~Gyr and 3~Gyr, for a wide variety of hypothesis on the progenitors \citep{2005A&A...441.1055G}. According to \citet{2011MNRAS.414.3231K}, an onset of the carbon production by the AGB stars refers to the time, when the galactic iron abundance grew to [Fe/H] $\simeq -1.5$.

With the SNeII + HNe model, \citet{2011MNRAS.414.3231K} predicted the close-to-solar C/Fe abundance ratio at low metallicities, $-4 \le$ [Fe/H] $\le -3$, but the [C/Fe] ratio becomes as large as 0.9, when including the yields of rotating massive stars. At [Fe/H] $\simeq -1$, [C/Fe] reaches 0.13, 
if all the sources, AGB + SNeIa + SNeII + HNe, are included, 
and [C/Fe] decreases down to $-$0.19, when excluding the AGB yields.

A role of different carbon sources at different metallicities can be investigated from confronting the observations with the chemical evolution models.
Studies in the literature demonstrate a variety of the observed trends of [C/Fe] versus [Fe/H].
For the sample of the $-3.2 <$ [Fe/H] $<-0.7$ dwarfs \citet{2004AA...414..931A} obtained enhanced C abundance relative to Fe, with [C/Fe] reducing towards higher metallicity from 0.45 to 0.25. They analysed the near infrared (IR) C~I lines under the local thermodynamic equilibrium (LTE) assumption. 
\citet{2006AA...458..899F} revised the C abundances of that stellar sample based on the non-local thermodynamic equilibrium (NLTE) line formation for C~I
and obtained the [C/Fe] ratio to be, on average, close to the solar one. 
\citet{2006MNRAS.367.1329R} studied the C~I lines in the visible spectral range in the sample of the thick disk and thin disk F-G dwarfs in the $-1.2 <$ [Fe/H] $< +0.2$ range. From the LTE analysis they 
obtained a supersolar [C/Fe] ratio of about 0.4 for the most metal-poor ([Fe/H] $< -0.4$) thick disk stars and a decline down to the subsolar values at higher metallicity. In the overlapping metallicity range, the thin disk stars reveal, on average, lower [C/Fe] ratios than the thick disk stars.  
In contrast, the study by \citet{2006MNRAS.367.1181B} based
on the [C~I] line at 8727\,\AA\ shows that [C/Fe] versus [Fe/H] trends for the thin and thick discs are totally merged and flat, with [C/Fe] $\simeq$ 0.1, for the subsolar metallicities, down to [Fe/H] = $-0.8$.

From observations of the CH bands in 83 subdwarfs 
\citet{1987PASP...99..335C} found [C/Fe] to be essentially constant and close to the solar value over the range $-2.5 <$ [Fe/H] $< -0.7$. However, they noted an upturn in the [C/Fe] values at [Fe/H] $< -2$. 
A remarkably flat [C/Fe] versus [Fe/H] relation, with [C/Fe] = +0.18, was found by
\citet{2005AA...430..655S} from observations of the CH bands for the sample of 'unmixed' giant stars in the range $-4 <$ [Fe/H] $< -2.5$.

The situation with understanding sources of the carbon production in the early Galaxy is complicated by a discovery of the Carbon-Enhanced Metal-Poor (CEMP) stars that show very high enhancements in carbon, with [C/Fe] up to 3~dex \citep[see recent papers of][]{2010AA...513A..72B,2013AA...552A.107S,2013ApJ...773...33I,2013AJ....146..132L,2013ApJ...778...56C,2013ApJ...778..146O,2014ApJ...797...21P, 2014ApJ...787....6K,2015AA...576A.118A, 2015ApJ...802L..22R,2015AJ....149...39A, 2015arXiv150507126S}.
It is still debated whether their carbon enhancements are inborn, or their atmospheres were polluted, most likely by accretion from an AGB binary companion. 
Recent study of \citet{2014MNRAS.441.1217S} confirmed a binarity of the CEMP-s stars that are additionally enhanced in barium. 
But new radial velocity data for the 15 CEMP-no stars, not enhanced in barium, were found to be inconsistent with the binary properties of the CEMP-s class, thereby strongly indicating a different physical origin of their carbon enhancements. 

Stellar carbon abundance determinations rely on various spectroscopic indicators. These are allowed and forbidden lines of C~I and lines of the molecular species CH, C$_2$, and CO. 
Suitable lines of C~I are located in the visible (4300 -- 7900\,\AA) and near-IR (7900 -- 20000\,\AA) spectral regions. All the allowed lines have close together excitation energies of the lower level, \Eexc, but different 
oscillator strengths, with smaller values for the visible than the near-IR lines. As a result, the visible C~I lines are much weaker than the near-IR ones.

The [C~I] forbidden line at 8727\,\AA\ can reliably be measured for
 the close-to-solar metallicity stars. For stars, not enhanced in carbon, the C~I visible lines can be used down to [Fe/H] = $-1.5$ and the IR lines down to [Fe/H] = $-2.5$. At the lower metallicity, [Fe/H] $< -2.5$, stellar carbon abundances are determined presumably from the CH molecular lines.

An use of different abundance indicators with classical 1D model atmospheres can produce systematic shifts and false metallicity trends.
 This is exactly the case in the \citet{1992AJ....104.1568T} study. They obtained, on average, 0.4~dex higher abundance from the atomic C~I compared with the molecular CH lines for the sample of metal-poor halo dwarfs. 
When dealing with a wide metallicity range, it is impossible to apply a common carbon abundance indicator. One needs, therefore, to investigate the possible sources of abundance discrepancy between the atomic and molecular lines. These can be (i) employing the homogeneous and plane-parallel (1D) model atmospheres, (ii) inadequate line-formation treatment based on the LTE assumption, (iii) and the uncertainties in stellar parameters.

First determinations of the carbon abundance based on a three-dimensional (3D), time-dependent and hydrodynamical model atmosphere were made by \citet{2005AA...431..693A} for the Sun. For the atomic lines, it was shown that the abundance differences between the 3D and 1D MARCS \citep{MARCS-ODF} models are small and amount to +0.01~dex, +0.01~dex, and $-0.02$~dex, on average, for the atomic C~I, molecular CH electronic and C$_2$ electronic lines, respectively. \citet{2009ARAA..47..481A} updated the calculations and obtained consistent abundances from the different species, with the mean log~$\epsilon_{\rm C}$ = $8.43\pm0.05$ for the 3D model and log~$\epsilon_{\rm C}$ = $8.42\pm0.05$ for the 1D MARCS model. 
\citet{2010AA...514A..92C} computed positive 3D-1D abundance corrections for the solar C~I lines, up to 0.1~dex for the stronger near-IR lines and $\le$0.03~dex for the fainter visible lines.

 For the stellar atmosphere parameters beyond the solar ones the 3D calculations were only performed for the C~I fictitious lines \citep{Dobrovolskas_Dis,2013A&A...559A.102D}. For the \Eexc\ = 6~eV lines at $\lambda$ = 8500~\AA\ in the \Teff/log~$g$/[M/H] = 5930/4.0/0, 5850/4.0/$-1$, and 5860/4.0/$-2$ models the 3D-1D abundance corrections do not exceed few 100-th in a logarithmic scale \citep[][Fig.~2.19]{Dobrovolskas_Dis}.

An influence of the 3D effects on abundance determination from the molecular CH lines was 
evaluated by \citet{2007AA...469..687C} and \citet{2011AA...529A.158H} for cool giants of various metallicity.  
\citet{2007AA...469..687C} showed that the 3D-1D abundance corrections for the CH lines 
at $\lambda \simeq$ 4360\,\AA\ are overall negative and they amount to $-$0.15~dex at solar metallicity and $-$0.8~dex at [M/H] = $-3$ in the models with \Teff\ = 5050~K and log~$g$ = 2.2.
\citet{2011AA...529A.158H} updated the \citet{2007AA...469..687C} calculations and obtained smaller 3D effects, namely 
3D$-$1D = 0.0, $-$0.18~dex, and $-$0.35~dex in the giant (log~$g$ = 2.2) models with [M/H] = 0 (\Teff\ = 5060~K), [M/H] = $-2$ (5050~K), and [M/H] = $-$3 (5100~K), respectively.

Thus, the discrepancy between the C~I and CH-based abundances found by \citet{1992AJ....104.1568T} in the classical 1D analysis is expected to be larger in 3D.


The NLTE methods for C~I were developed by \citet{1981ApJS...45..635V,1990KFNT....6...44S,1990AA...237..125S,1992PASJ...44..649T,2001AA...379..936P,2006AA...458..899F}, and \citet{2015MNRAS.446.3447L}.
\citet{1990AA...237..125S} showed that NLTE removes a discrepancy between visible and near-IR lines of C~I in the Sun, while the difference in LTE abundances between individual lines can be up to 0.37~dex, for example between C~I 9111\,\AA\ and 5380\,\AA. 
\citet{2005PASJ...57...65T} and \citet{2006AA...458..899F} investigated
the deviations from LTE for C~I in wide stellar parameter range and found that the NLTE corrections are overall negative and grow in absolute value towards lower carbon abundance. However, these two studies provide different results for hot very metal-poor models, for example the difference in NLTE correction for C~I 9094\,\AA\ amounts to 0.15~dex in the 6000/4.0/$-2$ model.
\citet{2006AA...458..899F} showed that an adequate line formation modelling for  C~I is important for stellar abundance determinations in Galactic chemical evolution studies, where the stellar sample covers a range of metallicities of more than 2~dex. Indeed, the LTE results of \citet{2004AA...414..931A} revealed an $\alpha$-element like behavior of carbon in the $-3 \le$ [Fe/H] $\le -0.7$ metallicity range, while close-to-solar C/Fe ratios were obtained by \citet{2006AA...458..899F} after correcting these data for the NLTE effects.   
The \citet{2015MNRAS.446.3447L} studied the NLTE effects for C~I in the galactic A-, F- and G-type supergiants and bright giants.

This study aims to construct a comprehensive model atom for C~I using the most up-to-date atomic data available so far and to determine the carbon abundance of the Sun and selected late-type stars with well-determined stellar parameters based on the NLTE line-formation for C~I in the classical 1D model atmospheres and high-resolution observed spectra. Our small sample covers the $-2.58 \le$ [Fe/H] $\le$ 0 range.
We also derive the carbon abundance from the molecular CH and C$_2$ lines and investigate the differences between the atomic and molecular lines, which can serve to characterise empirically how neglecting the 3D effects influences the derived abundances.

This paper is organized as follows. The model atom of carbon,
the adopted atomic data, and the departures from LTE for C~I depending on stellar parameters are described in Sect.\,\ref{Sect:atom}. 
Analysis of the atomic C~I and molecular CH and C$_2$ lines in the Sun is presented in Sect.\,\ref{Sect:Sun}. In Sect.\,\ref{Sect:Stars}, we determine the C abundance of the selected stars.
We summarise our conclusions in Sect.\,\ref{Sect:conclusions}.

\section{NLTE Line formation for C~I}\label{Sect:atom}

\begin{figure*}
\begin{minipage}{160mm}
\includegraphics[scale=0.6]{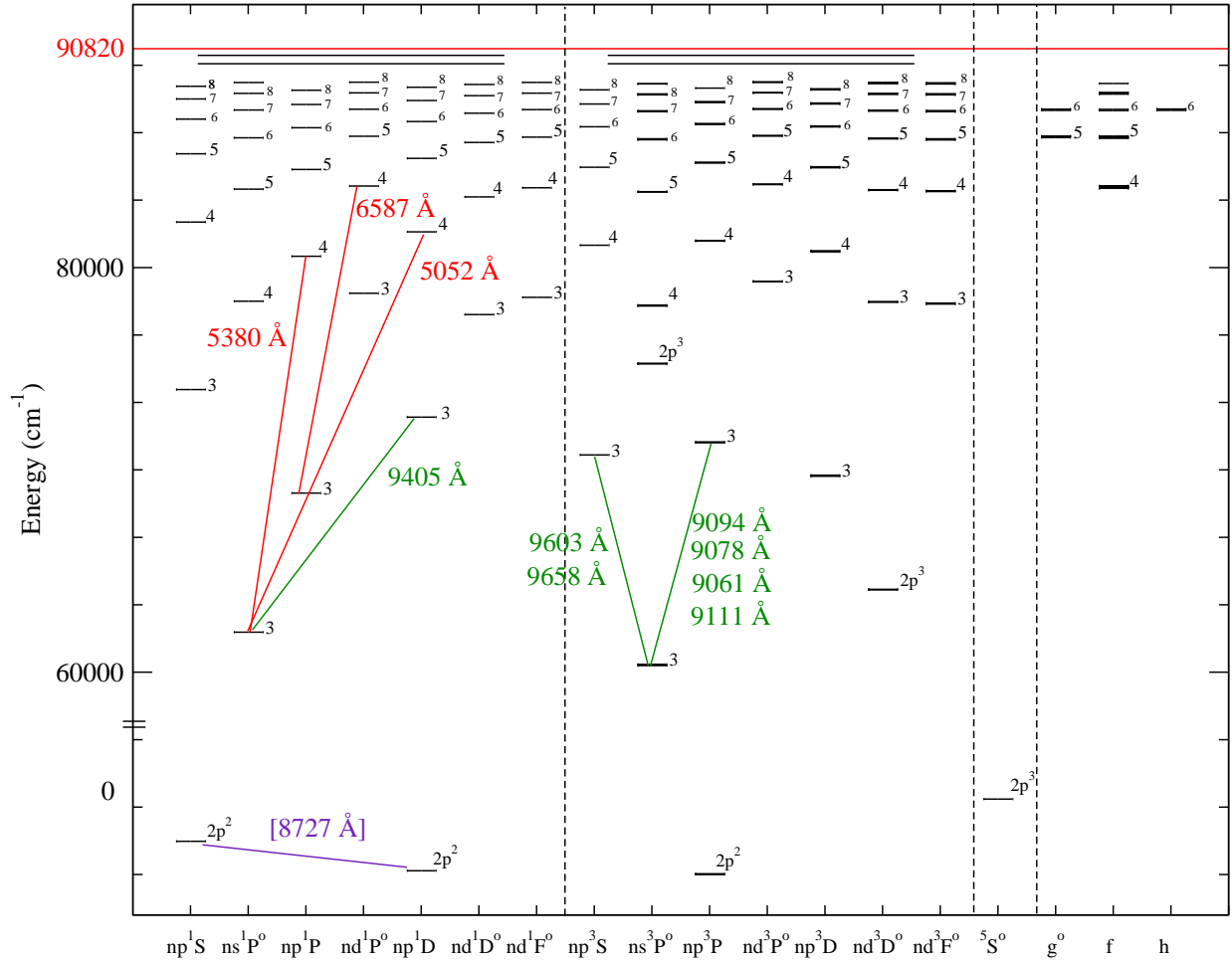}
\caption{Grotrian term diagram for neutral carbon. The solid lines indicate the seven transitions, where the investigated spectral lines arise. The $nf$, $ng$, $nh$ energy levels, where the Russell-Saunders coupling is broken, are shown on the right column.}
\label{Grot}
\end{minipage} 
\end{figure*}

\subsection{Model atom and atomic data}

{\bf Energy levels.} Model atom includes 208 energy levels of C~I up to $n$ = 10, $l$ = 4, nine lowest levels of C II, and the ground state of C~III. 
Most energy levels were taken from the NIST\footnote[1]{http://physics.nist.gov/PhysRefData/} database \citep{NIST08}.
They belong to singlet and triplet terms 
of the 2s$^2$2p~$nl$ ($n$ = 2-10,  $l$ = 0-2), 2s$^2$2p~$n$f ($n$ = 4-8), and 2s2p$^3$ electronic configurations and the 2s2p$^3$ quintet term.
In addition, the 2s$^2$2p~$nl$ ($nl$ = 5g, 6g, 6h) levels were taken 
from the Kurucz's database\footnote[2]{http://cfaku5.cfa.harvard.edu/atoms.html}.
To provide close collisional coupling of C~I to the continuum electron reservoir, the energy separation of the highest C~I levels from the ionization limit must be smaller than the mean kinetic energy of
electrons, i.e., 0.5~eV for atmospheres of solar temperature. In our model atom the energy gap between 
the uppermost levels of C~I and ionization limit is equal to 0.13~eV.
Fine structure splitting was included everywhere, up to $n$ = 8. 
All the states with $n$ = 9 and $n$ = 10 have close energies, and their populations must be in equilibrium to each other. Therefore, the levels of common parity were combined into the single superlevel. The Grotrian term diagram for our model atom is shown in Fig.\,\ref{Grot}. 

\noindent {\bf Radiative data.} 
Our model atom includes 1524 allowed bound-bound ($b-b$) transitions. 
Their transition probabilities were taken from the
NIST and VALD \citep{vald} databases, where available, and the Opacity Project database TOPbase\footnote[3]{http$//$legacy.gsfc.nasa.gov$/$topbase} \citep{1993BICDS..42...39C, 1989JPhB...22..389L,  1993AAS...99..179H}. Photo-ionization cross-sections for levels with $n \le 8$, $l \le 3$ were from TOPbase, and we adopted the hydrogen-like cross-sections for the two superlevels.

\noindent {\bf Collisional data.} 
All levels in our model atom are coupled via collisional excitation and ionisation by electrons and by neutral hydrogen atoms. Detailed electron-impact excitation cross-sections calculated in the close-coupling approximation using the
R-matrix method are available for more than 400 transitions from \citet{Reid1994}. For the remaining transitions we
use the impact parameter method \citep[IPM,][]{1962PPS....79.1105S} for the allowed transitions and assume that the effective collision strength $\Omega_{ij}$ = 1 for the forbidden transitions. Accurate data on inelastic collisions of carbon with neutral hydrogen atoms remain still undefined. We employed, therefore, the Drawin's formula \citep{1968ZPhy..211..404D}, as implemented by \citet{1984AA...130..319S}. The efficiency of C+H collisions is treated as a free parameter in our attempt to achieve consistent element abundances derived from the visible and near-IR lines in the Sun and selected stars.
For each object, the calculations were performed with a scaling factor of 
 $S_{\rm H}$ = 0, 0.1, 0.3, and 1. Hydrogen collisions for the forbidden transitions were ignored.
Ionisation by electronic collisions was everywhere treated through the \citet{Seaton1962} classical path approximation. The \citet{1984AA...130..319S} formula, with $S_{\rm H}$ was applied to calculate the C+H ionisation rates.

Our model atom is similar to that of \citet{2006AA...458..899F} except applying in this study detailed electron-impact excitation cross-sections from \citet{Reid1994}.

\subsection{Method of calculations}

To solve the radiative transfer and statistical equilibrium equations, we used the code DETAIL \citep{detail} based on the accelerated $\Lambda$-iteration method \citep{rh91}. The opacity package was improved as described in \citet{mash_fe}. 
The departure coefficients, $b_{\rm{i}}$ = $n_{\rm{NLTE}}$ / $n_{\rm{LTE}}$, were then used to compute
synthetic line profiles via the SIU (Spectrum Investigation Utility) code \citep{Reetz}.
Here, $n_{\rm{NLTE}}$ and $n_{\rm{LTE}}$ are the statistical equilibrium and thermal (Saha-Boltzmann) number densities, respectively. 
 
The calculations were performed with using plane-parallel, homogeneous
 (1D) MARCS model atmospheres \citep{2008AA...486..951G}. 

\subsection{Departures from LTE depending on carbon abundance and surface gravity}

\begin{figure}
\begin{center}
\includegraphics[scale=0.6]{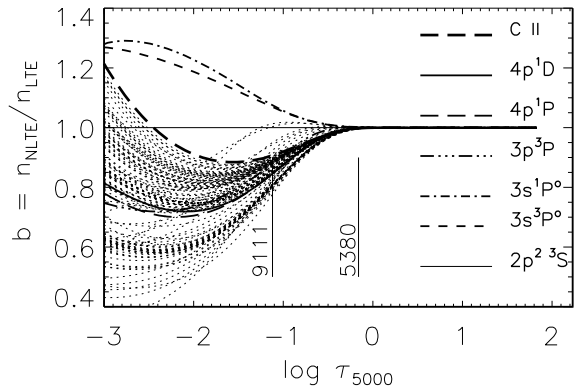}
\caption{Departure coefficients for the C~I levels and the C~II ground state as a function of $\log \tau_{5000}$ in the solar MARCS 5777/4.44/0.00 model atmosphere. The vertical lines mark the line-center formation depths for C~I 9111\,\AA\ and C~I 5380\,\AA.}
\label{BF}
\end{center}
\end{figure}

Figure\,\ref{BF} shows the departure coefficients for the selected levels 
in the solar 5777/4.44/0 model atmosphere. 
Neutral carbon C~I dominates the element number density over all atmospheric depths. 
Thus, no process seems to affect the C~I ground-state and low-excitation level populations significantly, and they keep their thermodynamic 
equilibrium values.
The levels 3s$^{1}$P$^{\circ}$ and 3s$^{3}$P$^{\circ}$ are overpopulated outward log$\tau = -0.2$ due to the ultraviolet radiative pumping in the transitions 2p$^{2}$ $^{1}$S -- 3s$^{1}$P$^{\circ}$ (2479\,\AA), 2p$^{2}$ $^{1}$D -- 3s$^{1}$P$^{\circ}$ (1931\,\AA), and 2p$^{2}$ $^{3}$P -- 3s$^{3}$P$^{\circ}$ (1657\,\AA).
Levels with \Eexc\ $>$ 8~eV are underpopulated outward log$\tau = -0.2$ due to spontaneous transitions to the lower levels.

The NLTE effects for a given spectral line can be understood from analysis of the departure coefficients at the line formation depths. The C~I lines used in abundance analysis are listed in Table\,\ref{tab0}. Here, we consider two lines with similar \Eexc, but different $gf$-values. The C~I 5380\,\AA\ (3s$^{1}$P$^{\circ}$ -- 4p$^{1}$P) line is weak, and it forms in the deep layers, around $\log \tau_{5000}$ = 0, where the departures from LTE are small. In contrast, C~I 9111\,\AA\ (3s$^{3}$P$^{\circ}$ -- 3p$^{3}$P) is strong, and its core forms around $\log \tau_{5000} = -1$, where the departure coefficient of the lower level, $b$(3s$^{3}$P$^{\circ}$) is larger than unity and $b$(3s$^{3}$P$^{\circ}$) $> b$(3p$^{3}$P). NLTE leads to strengthened lines of C~I and negative abundance corrections, $\Delta_{\rm NLTE}$ = log$\epsilon_{\rm NLTE}$ - log$\epsilon_{\rm LTE}$, with the stronger NLTE effects for the stronger lines.

\begin{figure}
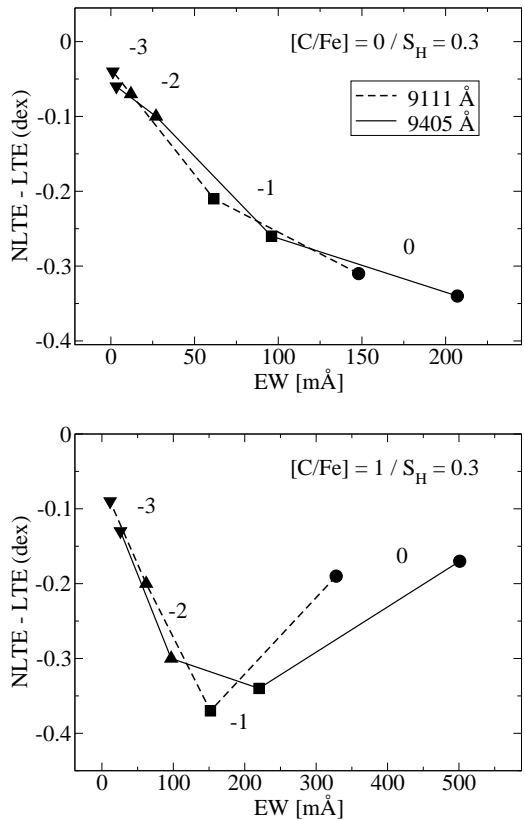

\begin{center}
\parbox{1\linewidth}{\includegraphics[scale=0.25]{Figure3.eps}\\
\centering}
\hspace{0.1\linewidth}
\hfill
\\[0ex]
\parbox{1\linewidth}{\includegraphics[scale=0.25]{Figure4.eps}\\
\centering}
\hfill
\end{center}
\caption{NLTE abundance corrections versus line equivalent width (EW) for C~I 9405~\AA\ (solid curve) and 9111~\AA\ (dashed curve) in the model atmospheres with $T_{\rm{eff}}$ = 6000~K, log$\, \textsl{g}$ = 4.0, and different metallicities of [M/H] = 0, $-1$, $-2$, and $-3$ (circles, squares, up-triangles, and down-triangles respectively). 
The NLTE calculations were performed with [C/Fe] = 0 (left panel) and [C/Fe] = 1 (right panel). Everywhere, $S_{\rm H}$  = 0.3 and $\xi_t$ = 1~\kms.}
\label{EW}
\end{figure}  

To evaluate the NLTE effects depending on the carbon abundance, we calculated the NLTE abundance corrections with the four model atmospheres having common $T_{\rm{eff}}$ = 6000~K and log$\, \textsl{g}$ = 4.0, but different metallicities, with [M/H] = 0, $-1$, $-2$, and $-3$. For each model the NLTE calculations were performed with the two different carbon abundances, [C/Fe] = 0 and [C/Fe] = 1. The $S_{\rm H}$ = 0.3 and a microturbulence velocity of $\xi_t$ = 1~\kms\ were employed. The obtained results are displayed in Fig.\,\ref{EW} for C~I 9405\,\AA\ and 9111\,\AA. 

\begin{figure*}
 \begin{minipage}{175mm}
\parbox{0.3\linewidth}{\includegraphics[scale=0.6]{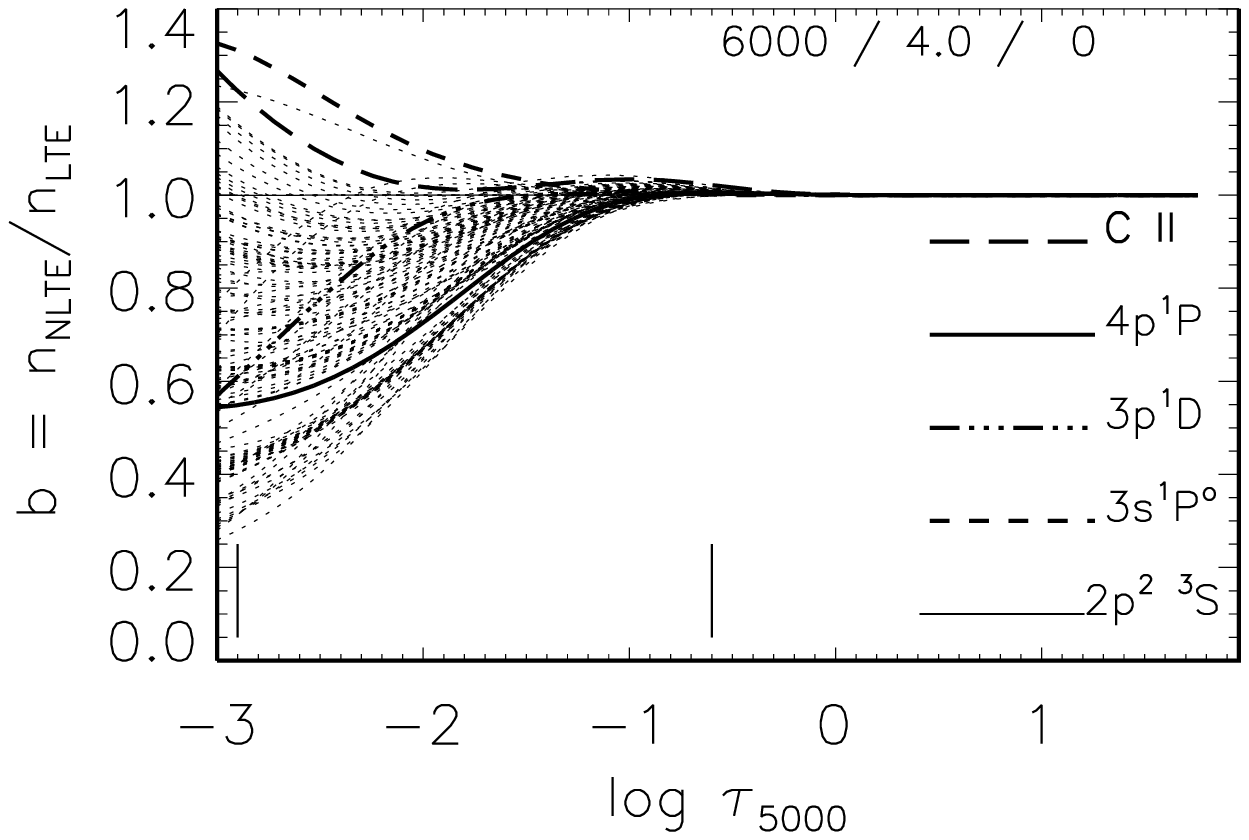}\\
\centering}
\hspace{0.2\linewidth}
\parbox{0.3\linewidth}{\includegraphics[scale=0.6]{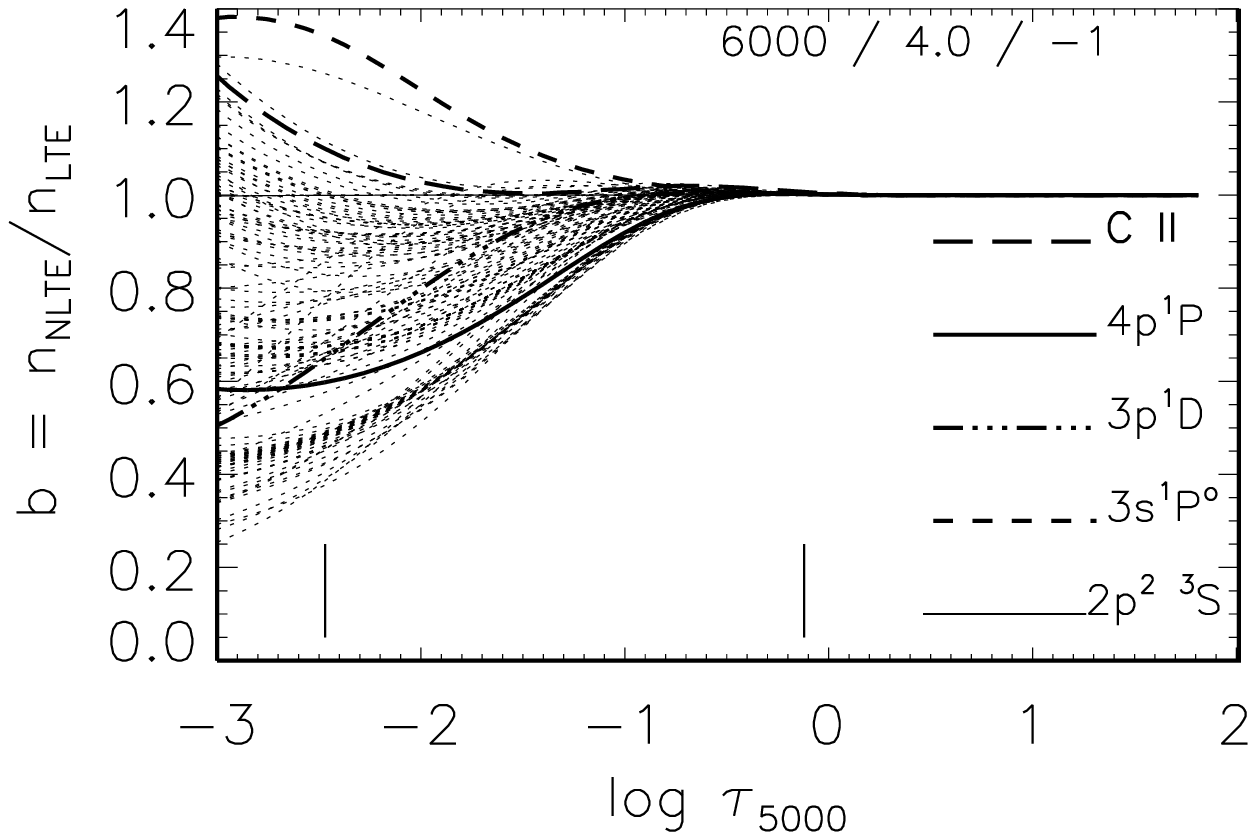}\\
\centering}
\hfill
\\[0ex]
\parbox{0.3\linewidth}{\includegraphics[scale=0.6]{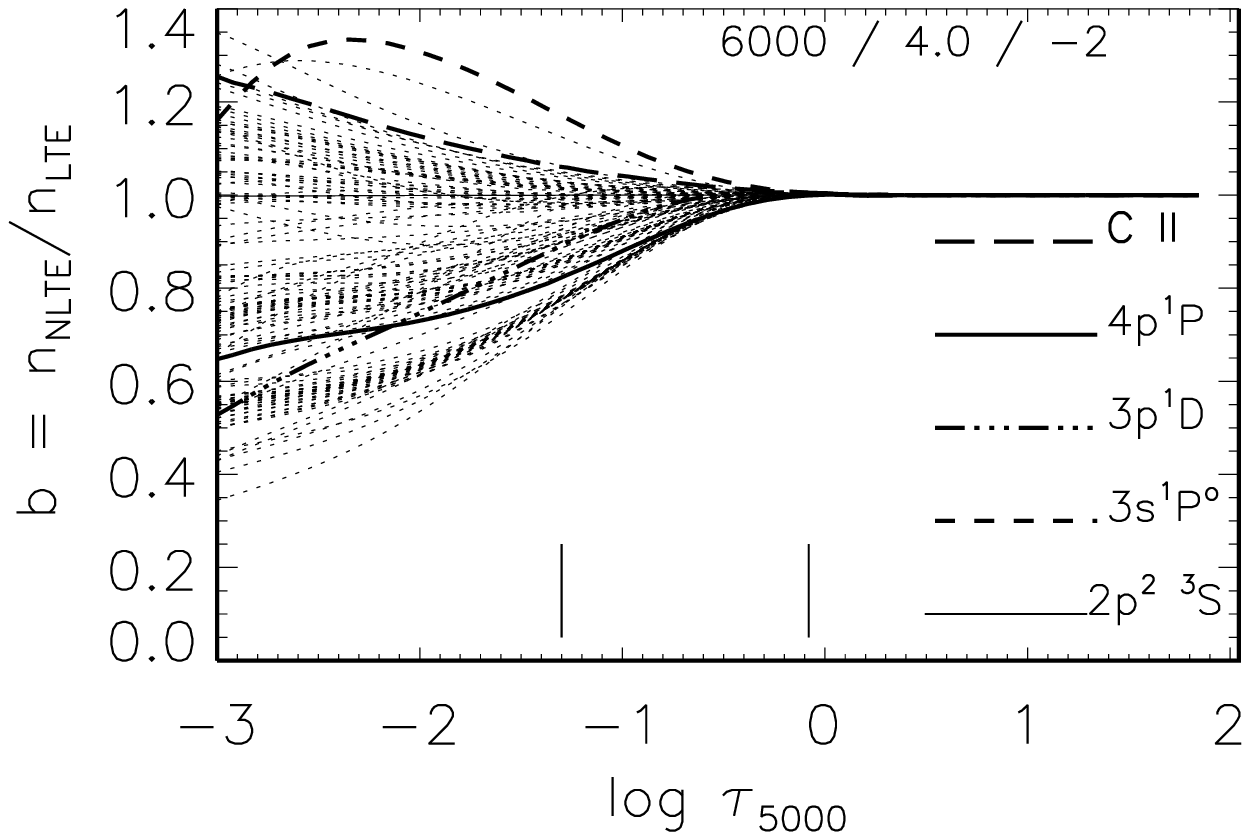}\\
\centering}
\hspace{0.2\linewidth}
\parbox{0.3\linewidth}{\includegraphics[scale=0.6]{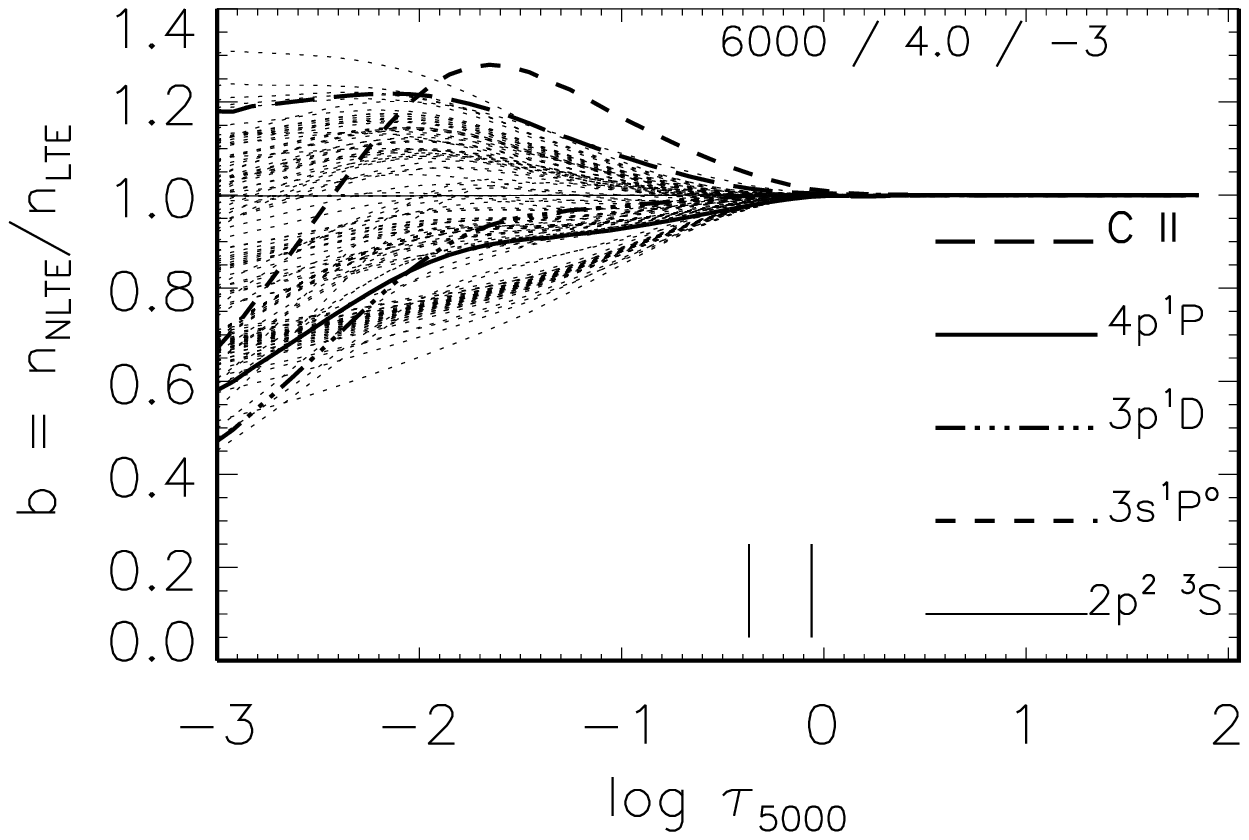}\\
\centering}
\hfill
\caption{Departure coefficients, $b$, for the C~I levels and the ground state of C~II as a function of log~$\tau_{5000}$
 in the four model atmospheres with different metal abundance: [M/H] = 0, $-1$, $-2$, and $-3$. Everywhere, $T_{\rm{eff}}$ = 6000~K, log$\, \textsl{g}$ = 4.0, [C/Fe] = 1, $\xi_t$ = 1~\kms. In each panel the two vertical lines indicate the formation region for C~I 9405\,\AA. }
\label{metal}
\end{minipage}
\end{figure*}

The NLTE effects for C~I depend on 
the element abundance. 
For [C/Fe] = 0 the NLTE abundance corrections reduce, in absolute value, towards lower equivalent width (element abundance) throughout the metallicity range from [M/H] = 0 to $-3$. 
To understand this, we consider the departure coefficients in Fig.\,\ref{metal}. 
On the one hand, the total number of free electrons, acting as the source of thermalisation, decreases with decreasing metallicity and the departures from LTE are amplified. Indeed, at a common optical depth the departure coefficients deviate from unity larger in the low than in the solar metallicity model.
 On the other hand, with metallicity (carbon abundance) decreasing, the line formation region moves to deeper layers, where the NLTE effects wane. Of the two competing effects the latter prevails, and the NLTE effects reduce toward lower metallicity.

A different behavior was found in the [C/Fe] = 1 case, namely the NLTE effects grow, when moving from [M/H] = 0 to $-1$, and then reduce towards lower EW.
The C~I 9405\,\AA\ and 9111\,\AA\ lines are very strong in the [M/H] = 0 model, with EW $\simeq$ 500\,m\AA\ and 330\,m\AA, respectively, which are mostly 
contributed from the broad line wings. The line wings form in deep atmospheric layers, where the departures from LTE are small. When moving to [M/H] = $-1$, 
a contribution of the line wings decreases, and the NLTE effects grow.
In the [M/H] = $-2$ and $-3$ models the NLTE effects decrease due to shifting the line formation regions in the deeper layers. 

\begin{figure} 
\begin{center}
\includegraphics[scale=0.25]{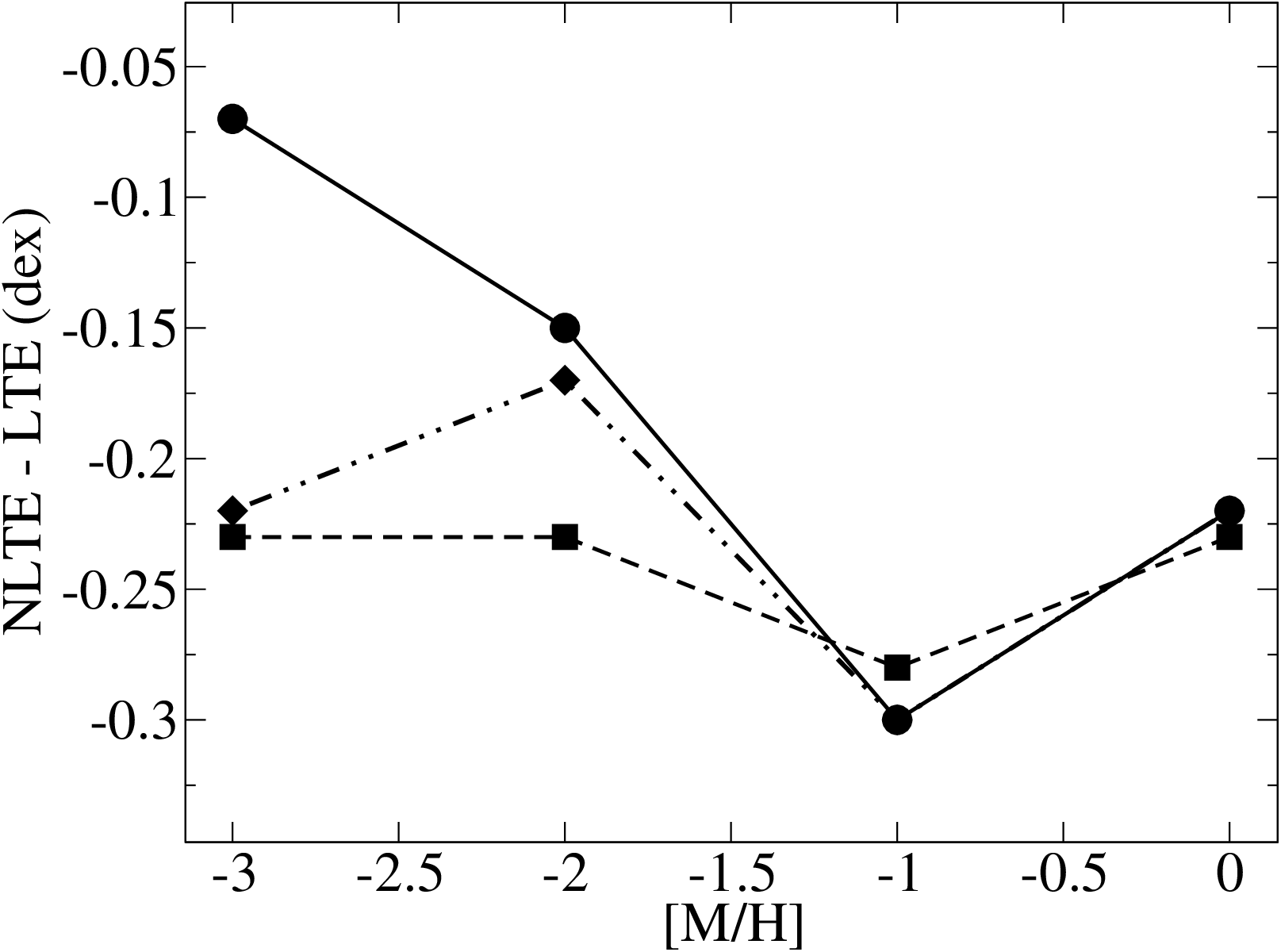}
\caption{NLTE abundance corrections depending on metallicity for C~I 9094\,\AA\ from this study (circles), \citet[][squares]{2006AA...458..899F}, and \citet[][rhombi]{2005PASJ...57...65T}. Everywhere, 
\Teff\ =6000~K, log~$g$ = 4.0, [C/Fe] = 0.4, $\xi_t$ = 1\,\kms, $S_{\rm H}$ = 1.}
\label{Comp}
\end{center}
\end{figure}

We compared our calculations with the NLTE results of  
\citet{2005PASJ...57...65T} and \citet{2006AA...458..899F}.
To be as close as possible to the NLTE method of \citet{2006AA...458..899F}, we employed the same collisional recipe as \citet{2006AA...458..899F} and $S_{\rm H}$ = 1.
Figure\,\ref{Comp} shows the NLTE abundance corrections depending on metallicity for the representative line at 9094\,\AA\ in the model atmospheres with common $T_{\rm{eff}}$ = 6000~K, log$\, \textsl{g}$ = 4.0, [C/Fe] = 0.4, and $\xi_t$ = 1~\kms. All the three studies give consistent results for the [M/H] = 0 and $-1$ models, where the NLTE effects are the strongest. At [M/H] = $-2$ our data agree well with \citet{2005PASJ...57...65T}, while a discrepancy of 0.08~dex in $\Delta_{\rm NLTE}$ was obtained with \citet{2006AA...458..899F}. 
The NLTE correction computed by \citet{2006AA...458..899F} for the [M/H] = $-3$ model is as large as that for [M/H] = $-2$, although the C~I 9094\,\AA\ line is much weaker in the [M/H] = $-3$ (EW = 8 m\AA\ ) than [M/H] = $-2$ (EW =
50.2 m\AA\ ) model. In our calculations $\Delta_{\rm NLTE} = -0.07$~dex. It is worth noting that spectral lines of EW  = 8 m\AA\ are too weak for accurate measurements, and, thus, the theoretical predictions cannot be checked with the observations.

\section{Solar carbon abundance }\label{Sect:Sun}
\subsection{Atomic C~I lines}

\begin{figure}
\begin{center}
\parbox{1\linewidth}{\includegraphics[scale=0.55]{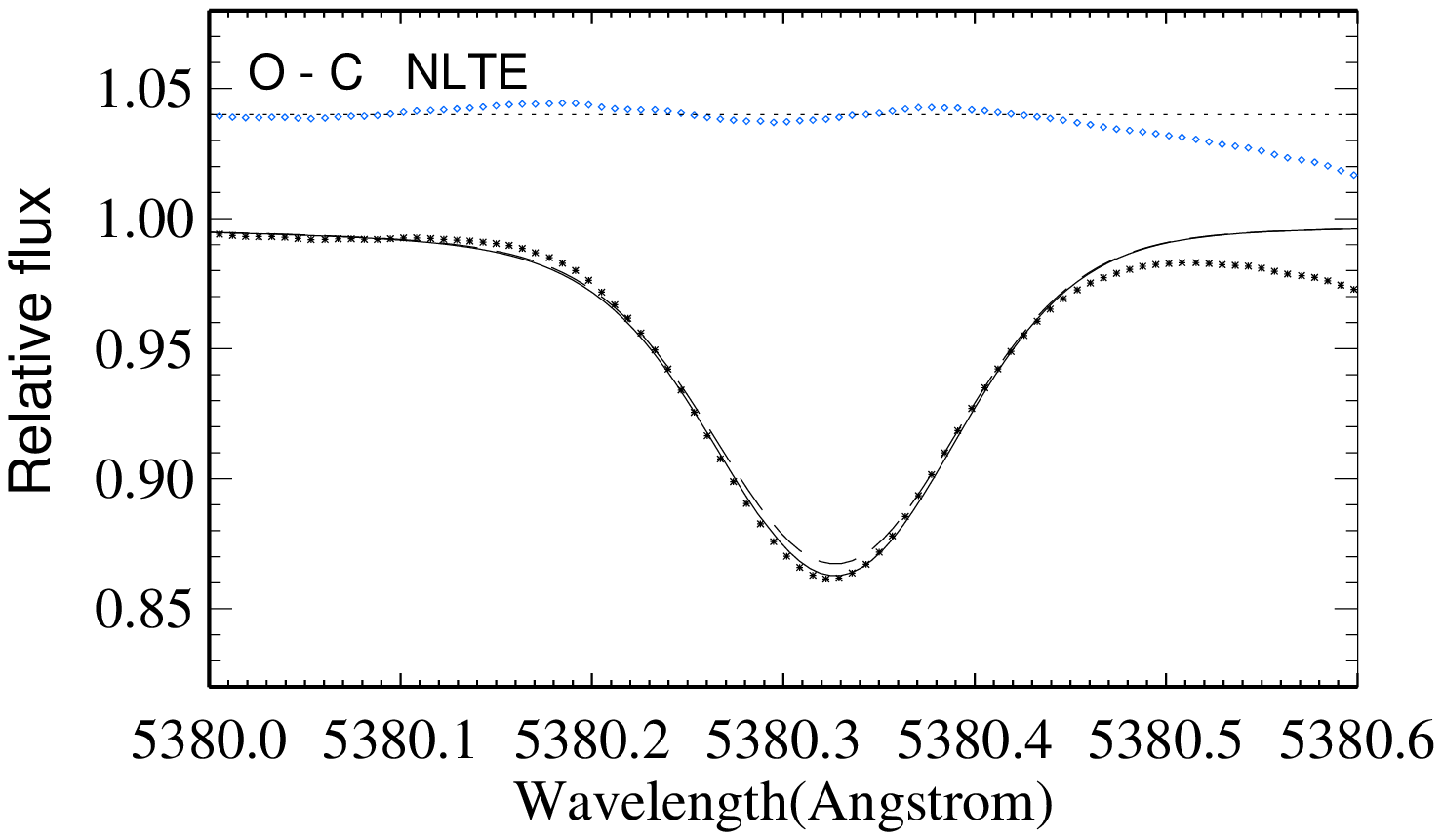}\\
\centering}
\hspace{0.1\linewidth}
\hfill
\\[0ex]
\parbox{1\linewidth}{\includegraphics[scale=0.55]{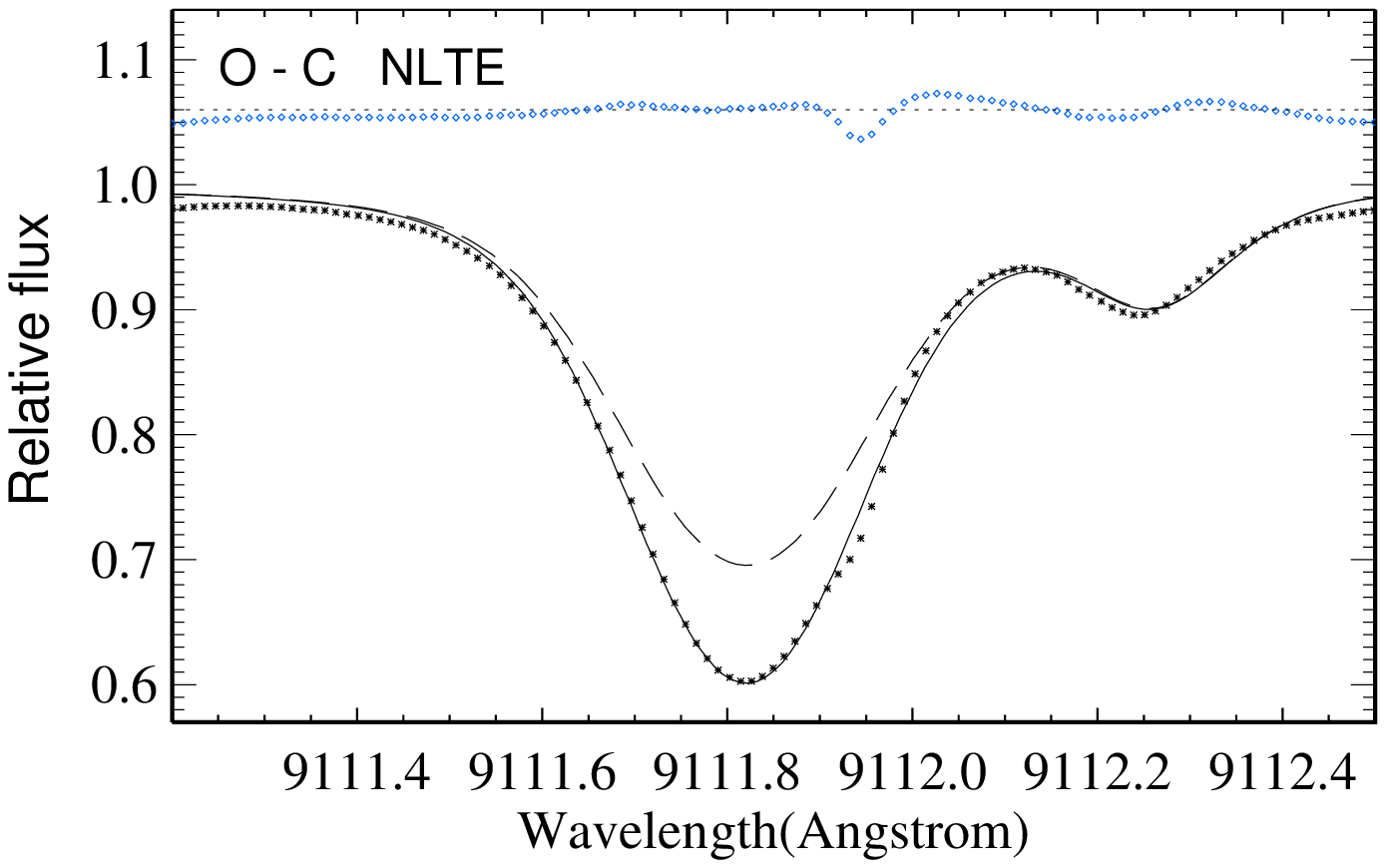}\\
\centering}
\hfill
\end{center}
\caption{Best NLTE fits (continuous curve) of the solar C~I 5380\,\AA\ and 9111\,\AA\
lines (asterisk). For each line, the LTE profile (dashed curve) was computed with the carbon abundance
obtained from the NLTE analysis. The differences between observed and calculated NLTE spectra, (O - C), are shown in the upper parts of the panels.
}
\label{sun}
\end{figure}

As a first application of the treated model atom, we derived the solar carbon abundance from lines of C~I. For comparison, the element abundance was also determined from the molecular CH lines. 
The solar flux observations were
taken from the Kitt Peak Solar Atlas \citep{1984sfat.book.....K}. 
We used the MARCS model atmosphere 5777/4.44/0 and a depth-independent microturbulence of 0.9\,\kms. 
The element abundance was determined from line profile fitting. As a rule, the uncertainty in
fitting the observed profile is
less than 0.02~dex for weak lines and 0.03~dex for strong lines. Our synthetic flux profiles were convolved with a profile that combines a rotational broadening of 1.8~\kms\ and broadening by macroturbulence with a radial-tangential profile. The most probable macroturbulence velocity V$_{mac}$ was varied between 2\,\kms\ and 4\,\kms\ for different lines of C~I and CH. Quality of the fits is illustrated in Fig.\,\ref{sun} for two lines of C~I. The C~I 5380~\AA\ profile beyond 5380.45~\AA\ is contributed from an unknown blend. Weak telluric line at 9111.95~\AA\ affects C~I 9111~\AA, however, the difference between observed and calculated NLTE spectra, (O - C), does not exceed 0.4~\%\ for the remaining profile.

\begin{table*}
\begin{minipage}{150mm}
   \begin{normalsize}
   \caption{Lines of C~I and solar carbon abundance from atomic lines}
        \label{tab0}
        \begin{tabular}{cccccccccc} \hline  
  \hline                                                       
$\lambda$, \AA    & Transition   & Mult.  & log~$gf$ & \Eexc       & LTE  &   \multicolumn{4}{c}{NLTE, $S_{\rm H}$ = }       \\ 
\cline{7-10}
&              &        &          &       &       & 0.0          &   0.1        & 0.3         & 1.0        \\\hline
8727.139          &2p$^{2}$ $^{1}$D -- 2p$^{2}$ $^{1}$S        &     & $-$8.165 & 1.26    & 8.45    & 8.45   &   8.45    &  8.45   &  8.45         \\ \hline
4932.049          &3s$^{1}$P$_{1}^{\circ}$ -- 4p$^{1}$S$_{0}$  & 13  & $-$1.658 &7.69     & 8.48    & 8.45         &   8.45       &    8.45     &  8.45          \\ 
5052.167          &3s$^{1}$P$_{1}^{\circ}$ -- 4p$^{1}$D$_{2}$  & 12  & $-$1.303 &7.69     & 8.50    & 8.47         &   8.47       &    8.47     &  8.47         \\
5380.337          &3s$^{1}$P$_{1}^{\circ}$ -- 4p$^{1}$P$_{1}$  & 11  & $-$1.616 &7.69     & 8.46    & 8.43         &   8.43       &    8.43     &  8.43         \\
6587.610          &3p$^{1}$P$_{1}$  -- 4d$^{1}$P$_{1}^{\circ}$ & 22  & $-$1.003 &8.54     & 8.45    & 8.43         &   8.43       &    8.43     &  8.43         \\ \hline
Mean (vis)     &              &        &          &             & 8.47     &8.45          &   8.45       &    8.45     &  8.45         \\ 
$\sigma$          &              &        &          &             & 0.02     & 0.02         &   0.02       &    0.02     &   0.02         \\
8335.148          &3s$^{1}$P$_{1}^{\circ}$ -- 3p$^{1}$S$_{0}$  & 10  & $-$0.437 &7.69    & 8.54    & 8.40         &   8.40       &    8.43     &  8.46    \\
9078.288          &  3s$^{3}$P$_{1}^{\circ}$ -- 3p$^{3}$P$_{1}$  & 3   & $-$0.581 &7.48  & 8.65    & 8.41      &   8.44       &    8.46     &  8.51  \\ 
9094.834          &  3s$^{3}$P$_{2}^{\circ}$ -- 3p$^{3}$P$_{2}$  & 3   & 0.151    &7.49  & 8.72    & 8.34      &   8.37       &    8.40     &  8.46  \\ 
9111.809          &  3s$^{3}$P$_{2}^{\circ}$ -- 3p$^{3}$P$_{1}$  & 3   & $-$0.297 &7.49  & 8.67    & 8.34      &   8.37       &    8.40     &  8.46   \\
9405.730          &  3s$^{1}$P$_{1}^{\circ}$ -- 3p$^{1}$D$_{2}$  & 9   & 0.286    &7.69  & 8.79    & 8.34      &   8.37       &    8.40     &  8.47   \\
9658.431          &  3s$^{3}$P$_{2}^{\circ}$ -- 3p$^{3}$S$_{1}$  & 2   & $-$0.280 &7.49  & 8.71    & 8.37      &   8.40       &    8.43     &  8.48   \\
9061.433          &  3s$^{3}$P$_{1}^{\circ}$ -- 3p$^{3}$P$_{2}$  & 3   & $-$0.347 &7.48  &  -   & -  &   -   &  -   &             - \\
9062.492          &  3s$^{3}$P$_{0}^{\circ}$ -- 3p$^{3}$P$_{1}$  & 3   & $-$0.455 &7.48  & -    & -  &    -  &  -   &             - \\\hline
Mean (IR)      &              &        &          &             & 8.68      &8.37          &8.39          &8.42    &8.47   \\ 
$\sigma$          &              &        &          &             &  0.08     &0.03          &0.03           &0.02    &0.02  \\ \hline
Mean            &              &        &          &             & 8.59      &8.41          &8.42          &8.43    &8.46   \\
$\sigma$          &              &        &          &             &  0.12     &0.05          &0.04           &0.03    &0.02   \\\hline
\multicolumn{10}{l}{Mult.: the multiplet numbers accordingly to \citet{1972mtai.book.....M}.} \\
  \end{tabular}
\end{normalsize}
 \end{minipage}
\end{table*} 
 
For lines listed in Table\,\ref{tab0} we determined the element abundance under various line-formation assumptions.
In LTE, the abundance difference between the visible and near-IR lines of C~I, $\Delta\log\epsilon$(vis - IR), amounts to $-0.21$~dex. 
In line with the previous studies, we find that the [C~I] 8727\,\AA\ forbidden line does not suffer from the NLTE effects, because it arises in the transition between the metastable levels that keep the LTE populations. 
The NLTE corrections are small for the visible lines, independent of the applied collisional recipes, with $|\Delta_{\rm NLTE}| \le 0.03$~dex. For the IR lines, the departures from LTE are sensitive to a variation in collisional rates. For example, for different lines $\Delta_{\rm NLTE}$ ranges between $-0.14$\,dex and $-0.45$\,dex, when $S_{\rm H}$ = 0, and between $-0.08$\,dex and $-0.32$\,dex, when $S_{\rm H}$ = 1. 
Consistent within 0.03~dex and 0.02~dex NLTE abundances from the visible and near-IR lines were obtained for $S_{\rm H}$ = 0.3 and 1, respectively. We derived from the atomic lines a solar carbon abundance of log$\epsilon_{\rm C}$ = 8.43$\pm$0.03 ($S_{\rm H}$=0.3).

\begin{table}
   \begin{normalsize}
  \caption{Carbon abundances derived from solar lines of C$_2$}
        \label{tab2a} 
  \begin{tabular}{lrcc}\hline 
  \hline 
$\lambda$, \AA\ & log($gf$)  & \Eexc    &   log$\epsilon$(C) \\ \hline                                      
 4992.2750    &   0.288  &    0.802         &    \multirow{2}{*}{8.44} \\                                          
 4992.3035    &   0.281  &    0.802         &      \\  \hline                                                                                                                                             
 5033.7792    &   0.191  &    0.584         &    8.47  \\  \hline                                         
 5143.3240    &  $-$0.411  &    0.102         &    8.44  \\  \hline     
 5144.9149    &  $-$0.447  &    0.097         &    8.46   \\  \hline    
 5145.2255    &  $-$0.485  &    0.098         &    8.46   \\  \hline    
 5150.5448    &   0.026  &    0.328         &    \multirow{4}{*}{8.46} \\     
 5150.5542    &   0.038  &    0.328         &                 \\
 5150.6461    &   0.013  &    0.328         &               \\
 5150.6737    &   0.638  &    0.084         &         \\  \hline                                                                  
 5052.6161    &   0.153  &    0.489         &    \multirow{2}{*}{8.45 } \\     
 5052.6254    &   0.162  &    0.489         &     \\  \hline                                                            
 5086.3897    &   0.031  &    0.328         &    8.48 \\  \hline   
 5103.7231    &  $-$0.023  &    0.250         &    \multirow{2}{*}{8.47} \\      
 5103.7710    &  $-$0.038  &    0.250         &           \\ \hline                                                               
 5109.0921    &  $-$0.053  &    0.227         &    \multirow{3}{*}{8.47} \\     
 5109.1490    &  $-$0.068  &    0.228         &         \\
 5109.3030    &  $-$0.084  &    0.228         &              \\ \hline                                                            
 5135.5509    &   0.137  &    0.471         &    \multirow{3}{*}{8.45} \\      
 5135.5818    &   0.127  &    0.472         &          \\
 5135.6825    &   0.117  &    0.472         &          \\ \hline                                                                
 5073.4490    &   0.090  &    0.388         &    \multirow{3}{*}{8.45}\\      
 5073.4513    &   0.101  &    0.388         &       \\
 5073.5815    &   0.080  &    0.388         &       \\ \hline
 Mean         &         &                   & 8.46$\pm$0.02 \\  \hline      
  \end{tabular}
\end{normalsize}
\end{table}

 \begin{table}
  \caption{ Carbon abundances derived from solar lines of CH}
        \label{tab2b} 
  \begin{tabular}{lccc}\hline 
  \hline 
$\lambda$, \AA\ & log($gf$)  & \Eexc[eV] &   log$\epsilon_{C}$ \\ \hline 
 4218.710  &   $-$1.315  &  0.413   &   \multirow{2}{*}{8.37}     \\           
 4218.734  &   $-$1.337  &  0.413   &            \\  \hline                    
 4248.729  &   $-$1.467  &  0.191   &   \multirow{3}{*}{8.38}      \\          
 4248.937  &   $-$1.431  &  0.191   &             \\                           
 4248.952  &   $-$3.256  &  0.191   &            \\\hline                      
 4253.000  &   $-$1.506  &  0.523   &   \multirow{2}{*}{8.41}      \\          
 4253.206  &   $-$1.471  &  0.523   &             \\\hline                     
 4255.248  &   $-$1.461  &  0.157   &   \multirow{2}{*}{8.41}     \\           
 4255.248  &   $-$3.210  &  0.157   &             \\\hline                     
 4263.969  &   $-$1.575  &  0.459   &  8.36      \\\hline                      
 4274.133  &   $-$3.025  &  0.074   &   \multirow{2}{*}{8.38}     \\           
 4274.186  &   $-$1.563  &  0.074   &             \\\hline                     
 4356.355  &   $-$1.846  &  0.157   &   \multirow{2}{*}{8.40}       \\         
 4356.371  &   $-$1.455  &  1.109   &              \\\hline                    
 4356.594  &   $-$1.793  &  0.157   &   8.39       \\         \hline
 Mean      &             &          &   8.39$\pm$0.02  \\ \hline
  \end{tabular}
\end{table}

\subsection{Molecular C$_2$ and CH lines}

{\bf C$_2$ electronic lines.} 
We selected the 12 least blended lines from the C$_2$ Swan band in the 4992-5150\,\AA\ range.
They are listed in Table\,\ref{tab2a} along with the lower level excitation potentials and the transition probabilities taken from \citet{2013JQSRT.124...11B}.

{\bf CH electronic lines.} We used a dissociation energy of D$_0$(C$_2$) = 6.297~eV from \citet{1991CPL...178..425U}.
Our analysis is based on the lines from the (0, 0) and (1, 1) bands of CH A-X around
4300\,\AA. They are listed in Table\,\ref{tab2b}. 
The lower level excitation potentials and the transition probabilities
 were taken from \citet{2014AA...571A..47M}. The dissociation energy is D$_0$(CH) = 3.465~eV \citep{Huber}. 
Our test calculations showed that using the molecular parameters from \citet{1996AA...315..204J} results in higher abundances, by 0.00 to 0.04~dex for different CH lines, 
compared with the corresponding values based on the \citet{2014AA...571A..47M} data.

Abundances from the individual molecular C$_2$ and CH lines are presented in Tables\,\ref{tab2a} and \ref{tab2b}. It can be seen that lines of C$_2$ give consistent within 0.04~dex abundances, and 
the mean abundance, log~$\epsilon_{\rm C}$ = 8.46$\pm$0.02,
agrees well with the NLTE($S_{\rm H}$ = 0.3) abundance from the atomic C~I lines.
Abundances from different lines of CH are consistent within 0.05~dex, and the mean abundance, log~$\epsilon_{\rm C}$ = 8.39$\pm$0.02, agrees with that from the atomic lines. 

\subsection{Comparison with previous studies}

 When applying a common $S_{\rm H}$ of 0, we find the obtained solar mean abundance from lines of C~I (8.41$\pm$0.05) to be consistent within the error bars with the 1D and 3D results of \citet[][hereafter, AGSS09]{2009ARAA..47..481A}, namely log~$\epsilon_{\rm C}$ = 8.39$\pm$0.04 (1D) and 8.42$\pm$0.05 (3D), and the 3D abundance log~$\epsilon_{\rm C}$ = 8.446$\pm$0.121 calculated by \citet[][hereafter, CLB10]{2010AA...514A..92C}.
Abundance from the forbidden [C~I] 8727\,\AA\ line, log~$\epsilon_{C}$ = 8.45, is 0.07~dex and 0.10~dex higher compared with the corresponding values from AGSS09 and CLB10. 
In part, this is due to employing in this study a 0.03~dex lower $gf$-value. Indeed, we relied on log~$gf = -8.165$ from \citet{2006JPhB...39.2159F}, while AGSS09 and CLB10 adopted log~$gf = -8.136$ by \citet{1993AAS...99..179H}. Another source of the discrepancy is a different treatment of faint blending lines.
According to the \citet{K07} data, two lines of Fe~I contribute to the 8727\,\AA\ blend. These are Fe~I 8727.10\,\AA\ (\Eexc\ = 5.587~eV, log~$gf = -5.924$) and Fe~I 8727.13\,\AA\ (\Eexc\ = 4.186~eV, log~$gf = -4.262$). \citet{2010AA...514A..92C} treated Fe~I 8727.132\,\AA, using the higher log~$gf = -3.93$ from the older \citet{1993sssp.book.....K} calculations. A difference of $-0.33$~dex in $gf$-value of the blending Fe~I line makes a +0.02~dex difference in the carbon abundance derived from [C~I] 8727\,\AA. 
To understand a source of the remaining discrepancy, we inspected the forbidden line in the solar flux \citep{1984sfat.book.....K} and disk-center intensity \citep{BraultTesterman} spectra and obtained fully consistent abundances, with $\log\epsilon_{\rm C}$ = 8.45. It is worth noting, we measured EW([C~I] 8727) = 5.5\,m\AA\ in the disk-center intensity spectrum, while AGSS09 and CLB10 reported lower values of 5.3~m\,\AA\ and 5.1~m\,\AA, respectively.

 Our 1D results for the molecular lines agree well with the 1D and 3D data of AGSS09, log~$\epsilon_{\rm C}$ =  8.40$\pm$0.03 (CH) and 8.46$\pm$0.03 (C$_2$) in 1D and log~$\epsilon_{\rm C}$ = 8.43$\pm$0.03 (CH) and 8.46$\pm$0.03 (C$_2$) in 3D.

\section{Determination of carbon abundances of the selected stars}\label{Sect:Stars}

\subsection{Stellar sample, observations, and stellar parameters}

 \begin{table}
 \begin{center}
     \caption{Atmospheric parameters of the selected stars and sources of the data.}
        \label{tab4} 
  \begin{tabular}{lccccc}\hline 
  \hline 
 Star &\Teff & log~$g$    &  [Fe/H] & $\xi_t$ &  Ref.   \\ 
  &[K]  &(CGS)    &   & [\kms] &    \\\hline 
   \multicolumn{6}{c}{Reference stars}\\
 HD ~29907  &   5500    &  4.64      &  -1.55  &  0.6 &  1 \\
 HD ~59374  &   5850    &  4.38      &  -1.02  &  1.2 &  3 \\
 HD ~84937  &   6350    &  4.09      &  -2.08  &  1.7 &  2 \\
 HD ~94028  &   5970    &  4.33      &  -1.50  &  1.4 &  3 \\
 HD 103095 &   5130    &  4.66      &  -1.26  &  0.9 &  3 \\
 HD 122563 &   4600    &  1.60      &  -2.56  &  2.0 &  2 \\ 
 HD 140283 &   5780    &  3.70      &  -2.38  &  1.6 &  3 \\
BD$-$4$^\circ$3208 &6390&  4.08     &  -2.20  &  1.3 &  3 \\
BD$+$66$^\circ$268 &5300&  4.72     &  -2.06  &  0.7 &  3 \\\hline
  \end{tabular}
\begin{tablenotes}  
\item[a] {\bf Notes.} References: (1) \citet{2003AA...397..275M};
(2) \citet{mash_fe}; (3) \citet{2015arXiv150601621S}
\end{tablenotes}  
\end{center}
\end{table} 

We selected nine metal-poor stars in the $-2.56 \le$ [Fe/H] $\le -1.02$ metallicity range, for which the high-quality observed spectra and the well determined stellar parameters are available in the literature (Table\,\ref{tab4}). 
We describe briefly the sources of stellar effective temperatures and surface gravities.

{\bf Effective temperatures.}
For HD~122563 we adopted \Teff\ = 4600~K from \citet{mash_fe}, and it is consistent with the recent value based on the interferometric observations of \citet{2012AA...545A..17C}. 
The infrared flux method (IRFM) temperatures were adopted for HD~84937 \citep[as recommended by][]{mash_fe}, 
HD~59374, HD~94028, HD~103095, HD~140283, BD$-$4$^\circ$3208, and BD$+$66$^\circ$268, as recommended by \citet{2015arXiv150601621S} based on the data of \citet{Alonso1996irfm,GH2009AA...497..497G,Casagrande2010}, and \citet{Casagrande2011}.
For HD~29907 its effective temperature was determined from the Balmer line wing fits \citep{2003AA...397..275M}.

{\bf Surface gravities.}
For HD~29907, HD~84937, HD~122563, and HD~140283 we adopted the Hipparcos parallax-based surface gravities calculated by \citet{2003AA...397..275M,mash_fe}, and \citet{2015arXiv150601621S}.
For the remaining five stars, log~$g$ were derived by \citet{2015arXiv150601621S} from
the NLTE analysis of the Fe~I and Fe~II lines. 

Spectral observations for four program
stars were carried out with the UVES spectrograph at the 8-m VLT telescope
of the European Southern Observatory (ESO) and for five stars with the Shane 3-m telescope of the Lick
observatory using the Hamilton echelle-spectrograph.
Characteristics of the observed spectra are summarized in Table\,\ref{tab3}.
All spectra were obtained with a high spectral resolving power, R, and high S/N ratio (more than 100). 
For HD~84937, HD~122563, and HD~140283 their observed spectra were taken from the
ESO UVESPOP archive \citep{2003Msngr.114...10B}.
The star HD~29907 was observed within the project 67.D-0086A, as described by \citet{2003AA...397..275M}. Detailed description of the spectra observed at the Lick observatory is given by \citet{2015arXiv150601621S}.

\begin{table*}
 \begin{minipage}{130mm}
     \caption{Characteristics of observed spectra. }
        \label{tab3} 
  \begin{tabular}{@{}lcllcccc}\hline 
  \hline 
Object     &V$^1$        & Telescope/   & Observing run   &Spectral range & $R$  &   $S/N$ \\
       &  (mag)  & spectrograph &   &\AA\        &      &       \\ \hline 
HD~~29907  &   9.91  &  8m VLT2/UVES    & Apr.2001 &3300$-$7000    &80\,000 & 150 \\
HD~~59374  &   8.50  &  3m Shane/Hamilton&Jan.2011, Mar. 2012&3700$-$9300    &60\,000 & 100 \\
HD~~84937  &   8.28  &  8m VLT2/UVES    &ESO UVESPOP &  3300$-$9900    &80\,000 & 200  \\
HD~~94028  &   8.22  &  3m Shane/Hamilton&Jan.2011, Mar.2012&3700$-$9300    &60\,000 & 100 \\
HD~103095 &   6.45  &  3m Shane/Hamilton&Jan.2011, Mar.2012&3700$-$9300    &60\,000 & 100 \\
HD~122563 &   6.20  &  8m VLT2/UVES    & ESO UVESPOP&  3300$-$9900    &80\,000 & 200 \\
HD~140283 &   7.21  &  8m VLT2/UVES    &ESO UVESPOP &  3300$-$9900    &80\,000 & 200 \\
BD$-$4$^\circ$3208 &  9.99   &  3m Shane/Hamilton&Jan. 2011, Mar.2012&3700$-$9300    &60\,000 & 100 \\
BD$+$66$^\circ$268 &  9.91   &  3m Shane/Hamilton&Jan. 2011, Mar.2012&3700$-$9300    &60\,000 & 100 \\ \hline
  \end{tabular}
  \begin{tablenotes}  
\item[a] $^1$ V is a visual magnitude from the SIMBAD database. 
\end{tablenotes}  
   \end{minipage}
\end{table*}

\subsection{Analysis of carbon atomic and molecular lines}

In this section we derive the carbon abundances of the selected stars from the atomic C~I and molecular CH lines
using the same approach as for Sun (Sect.\,\ref{Sect:Sun}) and the same set of 
atomic and molecular data.

The Shane/Hamilton observed spectra are affected by fringes in the IR spectral range.
For the three stars, HD~59374, BD$-$4$^\circ$3208, and BD$+$66$^\circ$268, we applied the local continuum normalisation procedure and this made possible to analyse the IR lines of C~I. 
For C~I 9094\,\AA\ and 9111\,\AA\ in HD~103095 and HD~94028 we adopted equivalent widths from \citet{1992AJ....104.1568T}.

The NLTE calculations for C~I were performed using various $S_{\rm H}$. 
As expected, everywhere NLTE leads to stronger C~I lines and negative abundance corrections.
For different stars $\Delta_{\rm NLTE}$ varies between $-0.06$~dex and $-0.45$~dex for the IR lines and between 0.00~dex and $-0.06$~dex for the visible ones. 
The NLTE corrections decrease in absolute value towards lower effective temperature and metallicity. For example, for C~I 9094\,\AA\ in HD~59374 (5850/4.38/$-$1.02) $\Delta_{\rm NLTE}$ = $-0.34$~ dex ($S_{\rm H}$ = 0.3), and the corresponding value is $-$0.04~dex for the cooler star HD~103095 (5130/4.66/$-$1.26).

\begin{figure*}
\begin{minipage}{150mm}
\parbox{0.3\linewidth}{\includegraphics[scale=0.17]{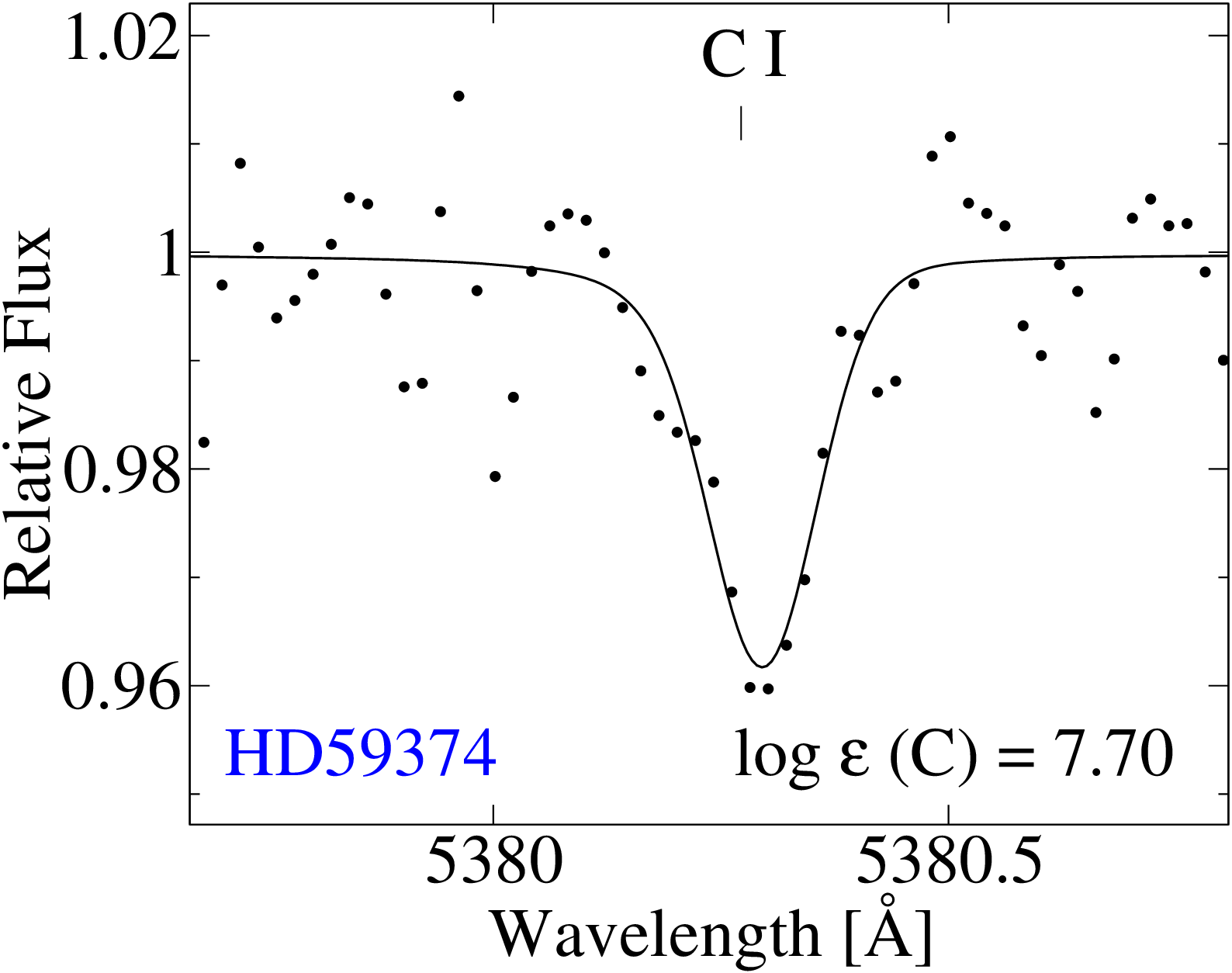}\\
\centering}
\parbox{0.3\linewidth}{\includegraphics[scale=0.17]{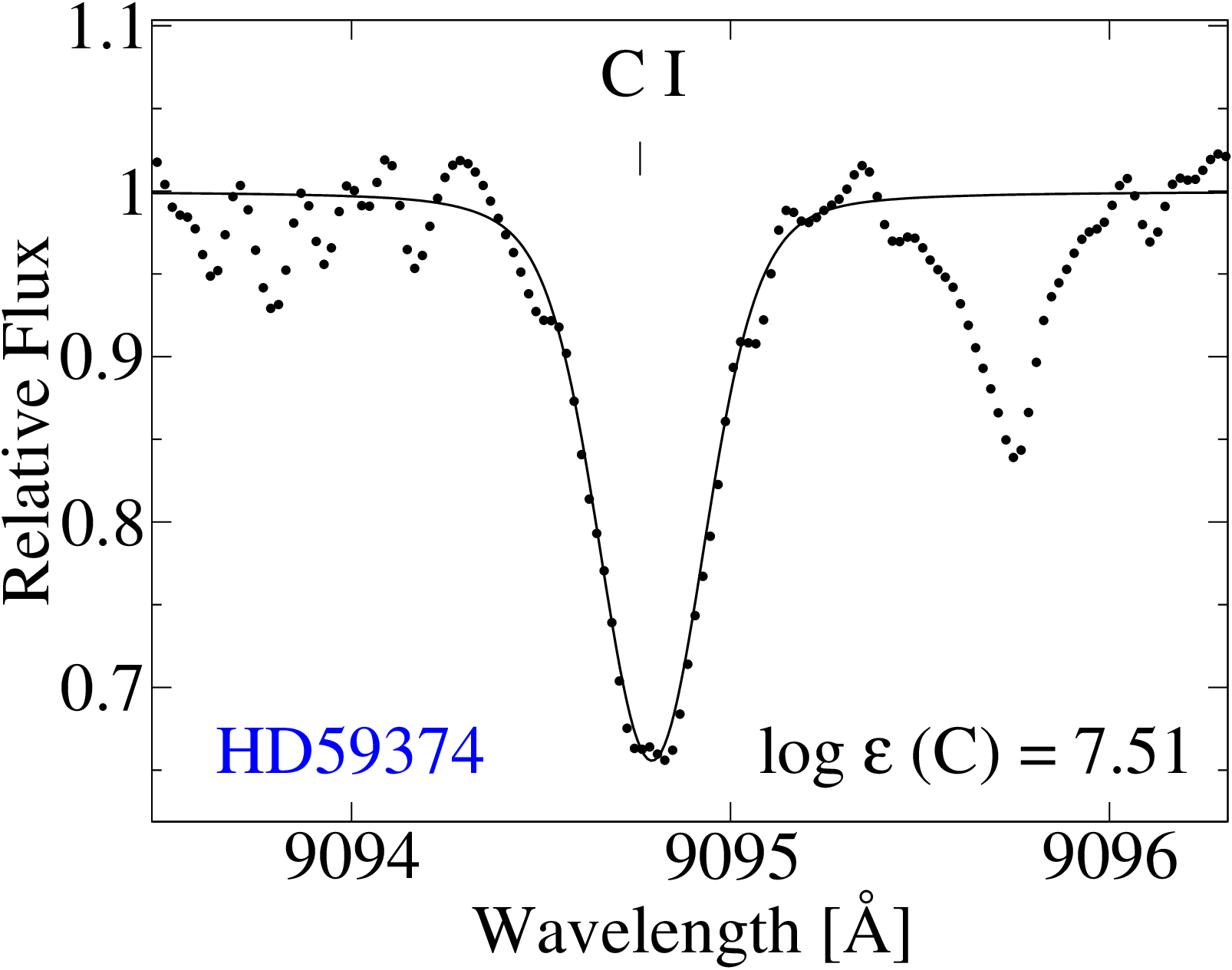}\\
\centering}
\parbox{0.3\linewidth}{\includegraphics[scale=0.17]{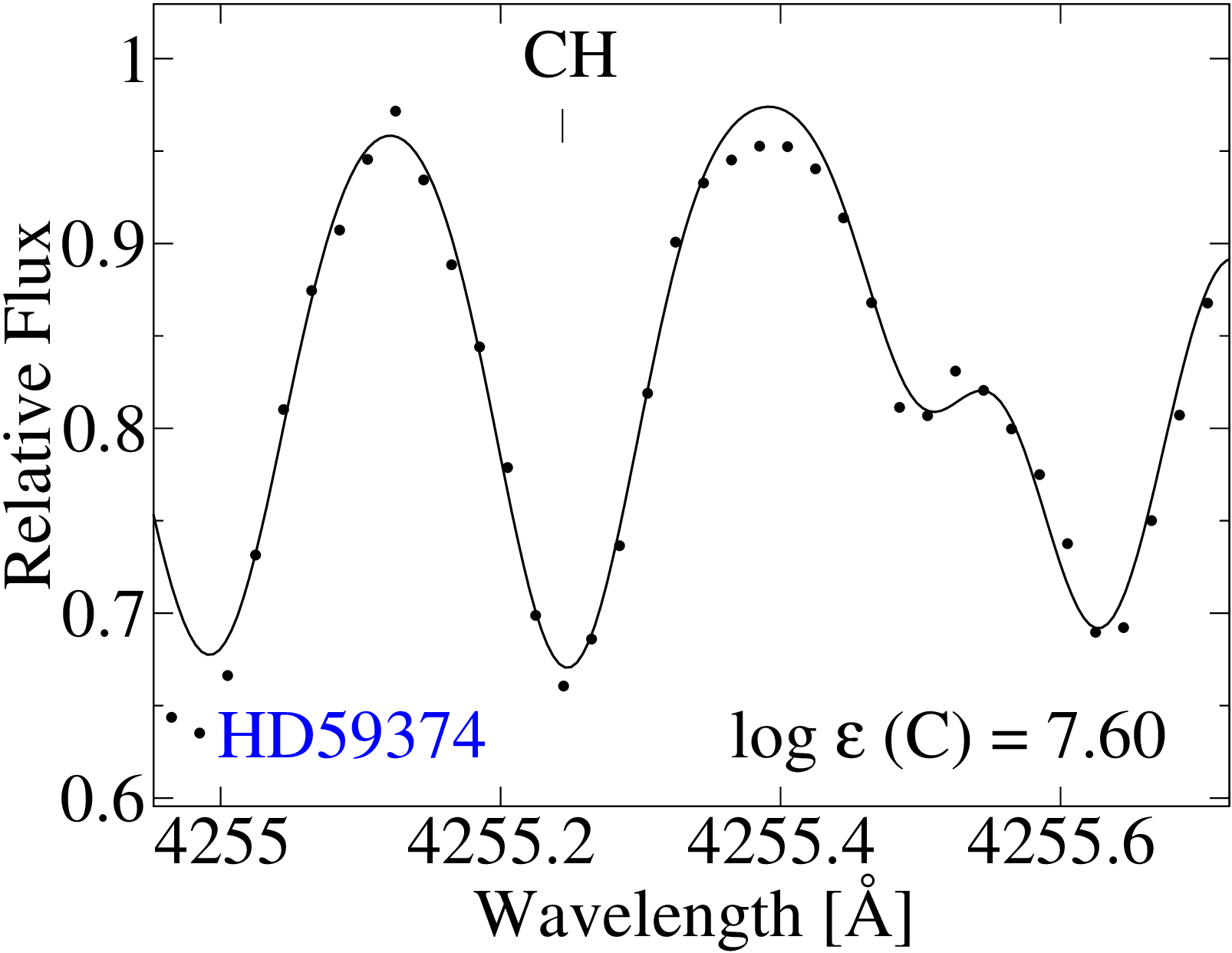}\\
\centering}
\hspace{1\linewidth}
\hfill
\\[0ex]
\parbox{0.3\linewidth}{\includegraphics[scale=0.17]{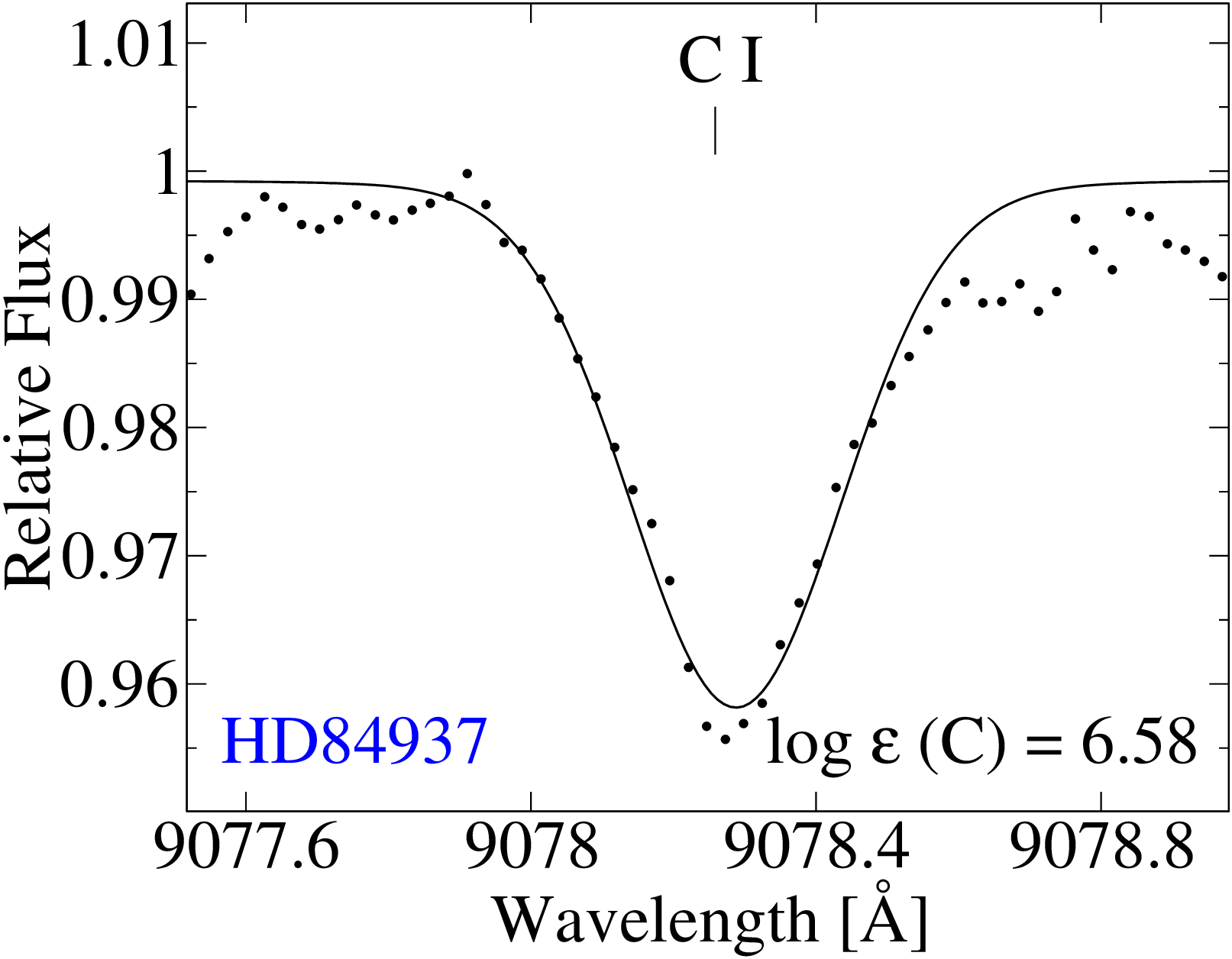}\\
\centering}
\parbox{0.3\linewidth}{\includegraphics[scale=0.17]{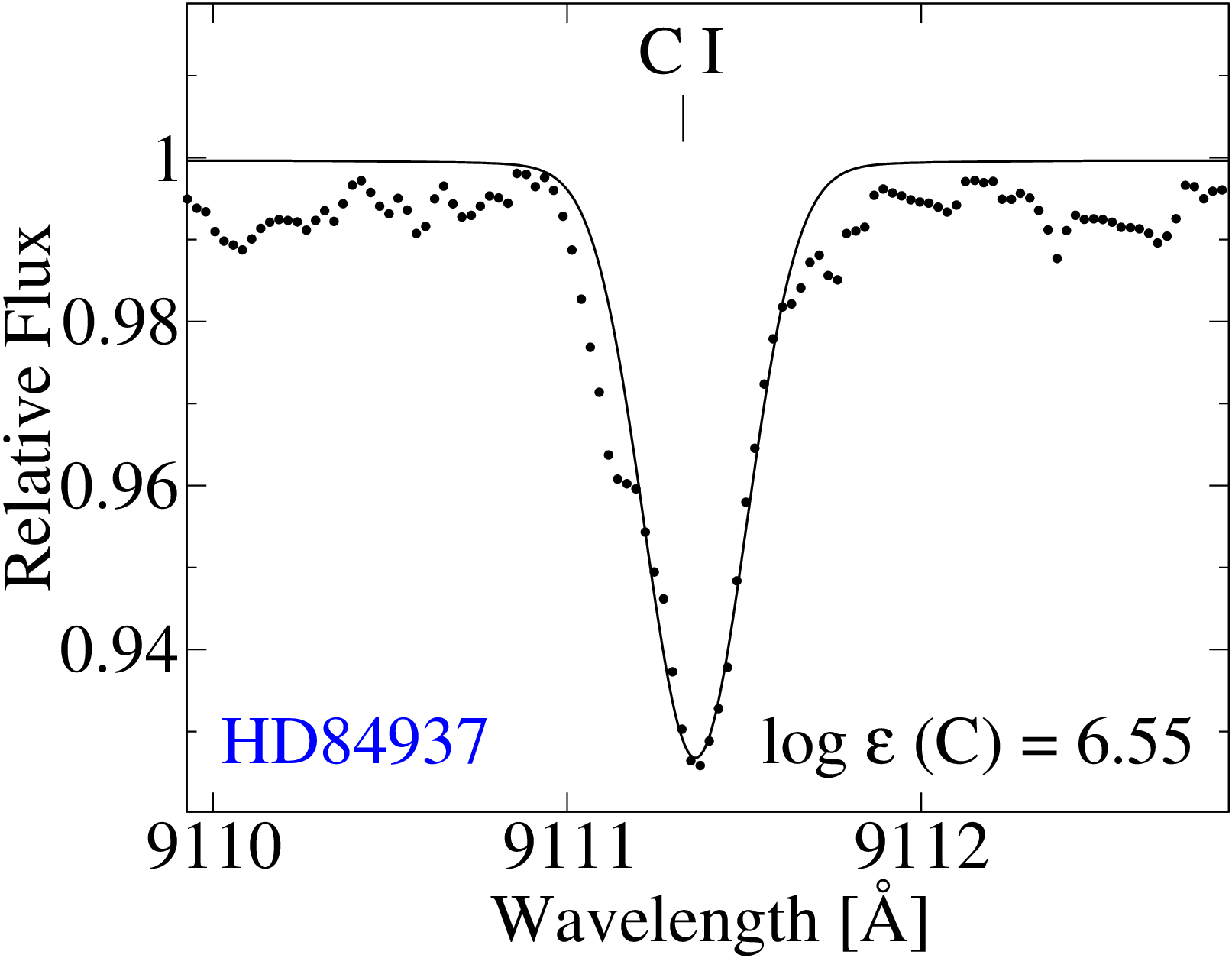}\\
\centering}
\parbox{0.3\linewidth}{\includegraphics[scale=0.17]{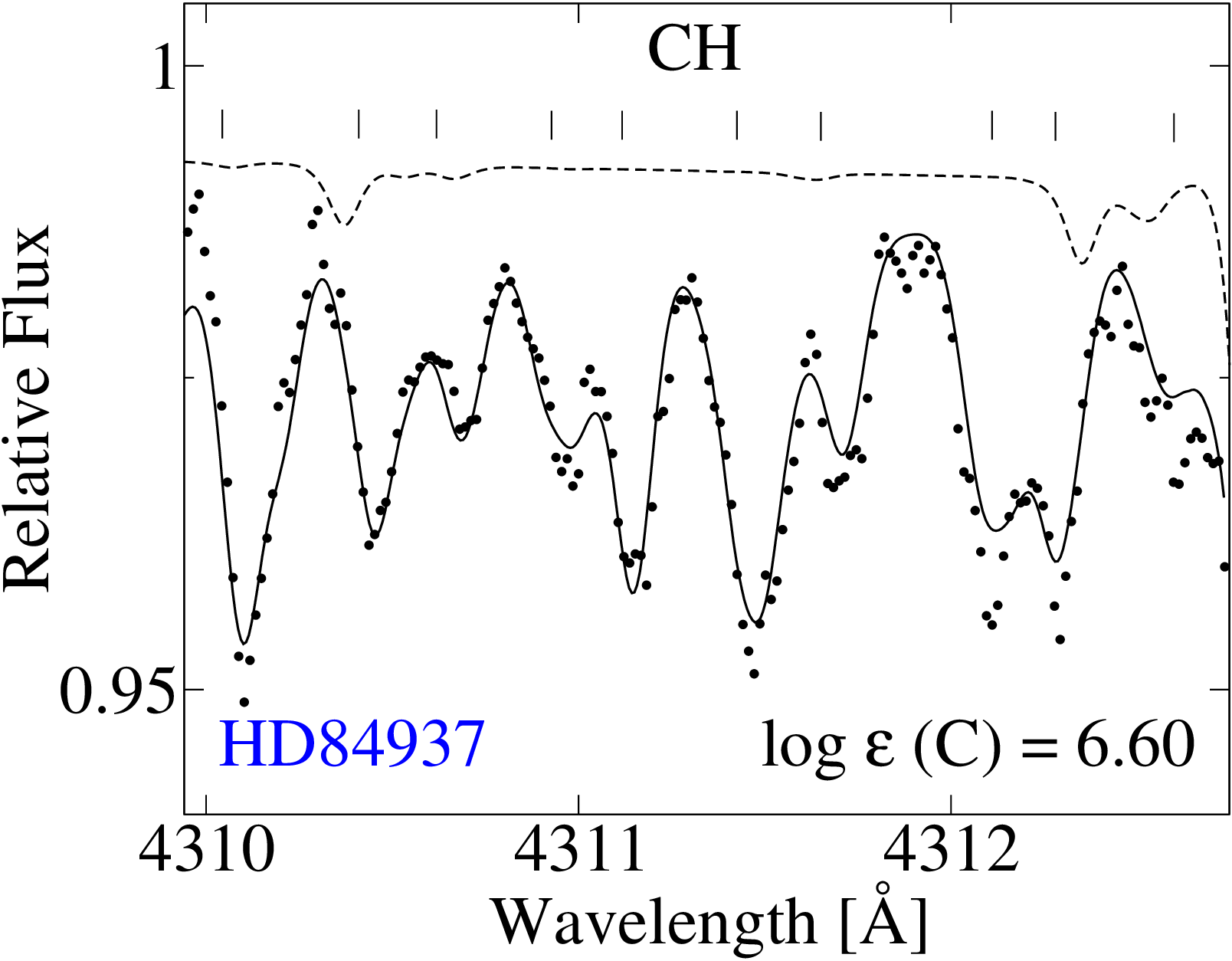}\\
\centering}
\hspace{1\linewidth}
\hfill
\\[0ex]
\caption{Best fits (continuum curves) of the selected atomic and molecular lines in 
HD~59374 (top row) and HD~84937 (bottom row).
The observed spectra are shown by bold dots. The dashed curve in the right bottom panel shows the
theoretical spectrum with no carbon in the atmosphere. 
 }
\label{pics}
\end{minipage}
\end{figure*} 

The best fits of some atomic and molecular lines in HD~59374 and HD~84937 are shown in Fig.\,\ref{pics}.

\begin{table}
   \begin{center}
   \begin{minipage}{150mm}
  \caption{LTE and NLTE abundances of the program stars. }
        \label{tab5} 
  \begin{tabular}{llccc}\hline 
  \hline                                                       
HD, BD & $\lambda$, \,\AA\ &  LTE           &                      \multicolumn{2}{c}{NLTE}                            \\ 
 &                 &                &   $S_{\rm H}$=0.1 &  $S_{\rm H}$=0.3  \\\hline 
29907 & 5052.144          & 6.82           & 6.79       & 6.79     \\\hline
59374 &4932.049          & 7.63           & 7.59        & 7.59      \\
&5052.167          & 7.75           & 7.70        & 7.71      \\
&5380.337          & 7.74           & 7.70        & 7.70      \\
&Mean(vis)         & 7.71           & 7.66        & 7.67     \\
&$\sigma$          & 0.07           & 0.06        & 0.07    \\
&9062.492          & 7.73           & 7.51        & 7.52      \\
&9078.288          & 7.80           & 7.60        & 7.62      \\
&9088.515          & 7.79           & 7.59        & 7.61      \\
&9094.834          & 7.88           & 7.51        & 7.54      \\
&9111.809          & 7.97           & 7.68        & 7.71       \\
&Mean(IR)          & 7.83           & 7.58        & 7.60     \\ 
&                  & 0.09           & 0.07        & 0.08     \\
&Mean(C~I)         & 7.79           & 7.61        & 7.63       \\
&$\sigma$          & 0.10           & 0.08        & 0.08     \\ \hline
84937& 8335.148          &    6.69        & 6.60       & 6.62   \\
&9062.492          &    6.68        & 6.59       & 6.61   \\
&9078.288          &    6.68        & 6.58       & 6.60   \\
&9111.809          &    6.67        & 6.55       & 6.57   \\
&Mean              &    6.68        & 6.58       & 6.60          \\
&$\sigma$          &    0.01        & 0.02       & 0.02      \\\hline
94028 &5052.167          & 7.19            & 7.14        & 7.16   \\
&9094.834$^1$  & 7.18            &6.90         &6.94     \\
&9111.809$^1$  & 7.19            &7.01         &7.04     \\
&Mean              & 7.19            & 7.02        &7.05      \\
&$\sigma$          & 0.01            & 0.12        & 0.11    \\\hline
103095 &9094.834$^1$  &6.92    &6.85      &6.88   \\
&9111.809$^1$  &7.14    &7.07      &7.10    \\
&Mean              &7.03    &6.96      &6.99    \\
&$\sigma$          &0.16    &0.16      &0.16     \\\hline
140283&8335.148          &6.43        &6.34        &6.35    \\
&9061.433          &6.57        &6.47        &6.48    \\
&9062.492          &6.52        &6.44        &6.45    \\
&9078.288          &6.51        &6.41        &6.42    \\
&9088.515          &6.55        &6.44        &6.45    \\
&Mean              &6.52        &6.42        &6.43      \\
&$\sigma$          &0.05        &0.05        &0.05       \\\hline
$-$4$^\circ$3208 &9061.433         & 6.34            &  6.21       &  6.22  \\
&9062.492         & 6.39            &  6.29       &  6.30  \\
&9111.809         & 6.44            &  6.29       &  6.30  \\
&Mean             & 6.39            &  6.26       &  6.27  \\
&$\sigma$         & 0.05            &  0.05       &  0.05 \\\hline
$+$66$^\circ$268& 9061.433         & 6.64                  &  6.56 &  6.58  \\
&9062.492         & 6.54                  &  6.47 &  6.48  \\
&Mean             & 6.59                  &  6.52 &  6.53  \\
&$\sigma$         & 0.07                  &  0.06 &  0.07  \\\hline
  \end{tabular}
      \begin{tablenotes}  
\item[a] $^1$ Equivalent widths from \citet{1992AJ....104.1568T}. 
\end{tablenotes}
   \end{minipage}
\end{center}
\end{table}

Carbon abundances obtained from the C~I lines are presented for all the stars in Table\,\ref{tab5} and from the molecular CH lines in Table\,\ref{tab6}. We comment below on the individual stars.

 

For the most metal-rich star of our sample, HD~59374, both visible and near-IR lines of C~I were measured, and we found the mean NLTE abundances from the two groups to be consistent, while $\Delta\log\epsilon$(vis - IR) = $-0.12$~dex in LTE. The least NLTE differences, $\Delta\log\epsilon$(vis - IR) = $+0.07$~dex and $+0.05$~dex, 
were obtained for $S_{\rm H}$ = 0.3 and 1.0, respectively.
This is in line with our empirical estimate of $S_{\rm H}$ from analysis of the solar lines (Sect.\,\ref{Sect:Sun}). 
When calculating the mean from the atomic and molecular lines in Table\,\ref{tab6}, we used everywhere the C~I-based abundances determined with $S_{\rm H}$ = 0.3.

For HD~94028 we analysed the single line in the visible region, C~I 5052\,\AA, and the two IR lines, with equivalent widths from \citet{1992AJ....104.1568T}.

Though the only atomic line, at 5052\,\AA, was measured for HD~29907, the derived abundance is consistent with that from the molecular lines, independent of the line-formation scenario. 
For two IR lines of C~I in HD~103095 we relied on equivalent widths from \citet{1992AJ....104.1568T} and obtained an abundance difference of more than 0.2~dex between the lines, independent of the line-formation scenario. 

For two stars we cannot inspect the difference in abundances between the atomic C~I and molecular CH lines. In BD$-$4$^\circ$3208 the molecular CH lines are too weak to be measured reliably, and only upper limit was estimated for the abundance from few individual lines (Table\,\ref{tab6}).
In contrast, in HD~122563 the atomic C~I lines cannot be measured, but stellar carbon abundance is reliably obtained from the molecular CH lines, with a standard deviation of 0.04~dex.

\begin{table*}
\begin{minipage}{160mm}
   \begin{normalsize}
  \caption{ Carbon abundances of the program stars from the CH A-X electronic lines.}
        \label{tab6} 
  \begin{tabular}{cccccccccc}\hline 
  \hline 
   HD/BD            &29907                 & 59374               & 84937              & 94028              &  103095            & 122563                &140283               &$-$4$^\circ$3208       & $+$66$^\circ$268       \\  \hline     
  $\lambda$, \AA\      &\multicolumn{9}{c}{log$\epsilon_{C}$ }   \\ \hline 
4362.4 - 4364.6&  6.77                  &   --                  &   6.58               &   7.12               & 7.07                 & --                      &   6.35                &   --                    &   6.34                   \\
4366.0 - 4367.0&  6.78                  &   --                  &   6.58               &   7.12               & 7.10                 & --                      &   6.32                &   --                    &   6.39                   \\
4310.0 - 4312.5&  6.73                  &    --                 &   6.60               &   6.99               & --                   &5.13                     &   6.35                &   --                    &   6.34                   \\
4313.4 - 4313.8&    --                  & --                    &   6.64               &   7.00               & --                   &5.13                     &   6.30                &   --                    &   6.39                  \\\hline
 4218.710      &  \multirow{2}{*}{6.71} & \multirow{2}{*}{7.48} &\multirow{2}{*}{6.62} &\multirow{2}{*}{7.08} & \multirow{2}{*}{--}  &\multirow{2}{*}{5.09}    &\multirow{2}{*}{6.35}  &\multirow{2}{*}{--}      &  \multirow{2}{*}{6.34}   \\
 4218.734      &                        &                       &                      &                      &                      &                         &                       &                         &                        \\\hline 
 4248.729      &  \multirow{3}{*}{6.74}  &\multirow{3}{*}{7.49}  &\multirow{3}{*}{-- } &\multirow{3}{*}{7.05}&\multirow{3}{*}{6.93} &\multirow{3}{*} {5.12}    &\multirow{3}{*}{6.38}  &\multirow{3}{*}{--}  &\multirow{3}{*}{6.34}   \\      
 4248.937      &                        &                       &                      &                      &                      &                         &                       &                         &                        \\
 4248.952      &                        &                       &                      &                      &                      &                         &                       &                         &                        \\\hline 
 4253.000      &  \multirow{2}{*}{6.79}  &\multirow{2}{*}{7.55}  &\multirow{2}{*}{-- }  &\multirow{2}{*}{--}  &\multirow{2}{*}{7.01} &\multirow{2}{*}{5.09}    &\multirow{2}{*}{6.35}  &\multirow{2}{*}{--}      &  \multirow{2}{*}{6.39}   \\
 4253.206      &                        &                       &                      &                      &                      &                         &                       &                         &                        \\\hline  
 4255.248      &  \multirow{2}{*}{6.74} & \multirow{2}{*}{7.60} &\multirow{2}{*}{6.67} &\multirow{2}{*}{7.06} &\multirow{2}{*}{6.93} &\multirow{2}{*}{5.15}    &\multirow{2}{*}{6.40}  &\multirow{2}{*}{$<$ 6.33}  &\multirow{2}{*}{6.34}   \\
 4255.248      &                        &                       &                      &                      &                      &                         &                       &                         &                        \\\hline 
 4263.969      &  \multirow{2}{*}{6.79} & \multirow{2}{*}{7.57} &\multirow{2}{*}{-- }  &\multirow{2}{*}{7.05} & \multirow{2}{*}{--}  &\multirow{2}{*}{5.01}    &\multirow{2}{*}{6.37}  &\multirow{2}{*}{--}&  \multirow{2}{*}{6.36}   \\
               &                        &                       &                      &                      &                      &                         &                       &                         &                        \\\hline 
 4274.133      &  \multirow{2}{*}{6.75} & \multirow{2}{*}{7.63} &\multirow{2}{*}{-- }  &\multirow{2}{*}{7.07} & \multirow{2}{*}{--}  &\multirow{2}{*}{5.09}    &\multirow{2}{*}{6.42}  &\multirow{2}{*}{--}      &  \multirow{2}{*}{6.36}   \\
 4274.186      &                        &                       &                      &                      &                      &                         &                       &                         &                        \\\hline  
 4356.355      &  \multirow{2}{*}{6.78} & \multirow{2}{*}{7.64} &       6.63           &       --             &\multirow{2}{*}{7.08} & \multirow{2}{*}{-- }    &   6.46                &   --                    &   6.37 \\
 4356.371      &                        &                       &       6.50           &       --             &                      &                         &   6.39                &   --                    &   6.37               \\\hline
 4356.594      &  6.76                  &   7.63                &      --              &      --              & 7.05                 &--                       &    --                 &    --                   &                    \\\hline
 Mean          &  6.76                  &   7.57                &  6.60                & 7.06                 & 7.02                 &   5.10                  &   6.37                &  $<$ 6.33                &   6.36                 \\
$\sigma$       &  0.03                  &   0.06                &  0.05                &  0.04              &   0.07               &   0.04                  &   0.04                &                         &   0.02                 \\ \hline \hline
C~I+CH  & 6.78    & 7.60    & 6.60    & 7.06    &   7.01     & 5.10      & 6.40  &  6.27  & 6.45   \\
$[$C/Fe$]$  & $-$0.10   & 0.19    & 0.25    &  0.13   &   $-$0.16    &  $-$0.71    & 0.35  &  0.04  & 0.08   \\\hline  
 \hline
  \end{tabular}
\end{normalsize}
\end{minipage}
\end{table*}

\begin{figure} 
\includegraphics[scale=0.3]{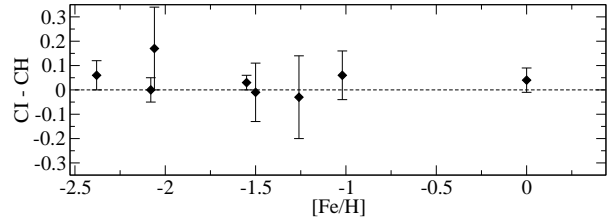}
\caption{Differences in abundances derived from the atomic C~I and molecular CH lines for the program stars.  }
\label{Teff}
\end{figure}

Figure\,\ref{Teff} displays the abundance differences between the atomic C~I and molecular CH lines for the Sun and seven metal-poor stars of our sample. Nowhere, $\Delta\log\epsilon$(C~I - CH) exceeds 0.07~dex and reveals any metallicity dependence. This suggests a minor influence of the 3D effects on abundance determinations from the molecular CH lines for metal-poor stars. The only outlier is BD$+$66$^\circ$268, with (C~I - CH) = 0.17~dex. 



We evaluated an influence of the uncertainties in stellar atmosphere parameters, namely 100~K in \Teff, 0.1~dex in log~$g$, and 0.1~dex in [Fe/H],
on the derived carbon abundances for five representative sets of \Teff/log~$g$/[Fe/H]. 
The atomic lines were represented by C~I 9094\,\AA\ and the molecular lines by the CH band 4310--4313\,\AA. The obtained results are presented in Table\,\ref{error}.

 \begin{table*}
 \begin{minipage}{150mm}
     \caption{Changes in the abundances derived from the C~I and CH lines caused by changes in stellar parameters.
     } \label{error}
  \begin{tabular}{lcccccccccc}\hline 
  \hline 
Models & \multicolumn{2}{c}{5100/4.5/$-$1}  & \multicolumn{2}{c}{5800/4.3/$-$1} & \multicolumn{2}{c}{5800/3.7/$-$2} & \multicolumn{2}{c}{6300/4.0/$-$2} & \multicolumn{2}{c}{4600/1.6/$-$2.5} \\
                          &   C~I     &   CH       &  C~I      &  CH       & C~I         &   CH       &    C~I       &   CH      &   C~I      &   CH      \\ \hline
$\Delta$\Teff=+100 K   &  $-$0.17  &   $-$0.01    &  $-$0.08  &  $+$0.12  & $-$0.06     &   $+$0.18  &    $-$0.05   &   $+$0.17 &   $-$0.11  &   $+$0.15 \\
$\Delta$log$g$=+0.1        &  $+$0.04  &   $+$0.01    &  $+$0.03  &  $-$0.01  & $+$0.04     &   $-$0.03  &    $+$0.03   &   $-$0.04 &   $+$0.05  &   $-$0.03 \\
$\Delta$ $[M/H]$=+0.1      &  $+$0.03  &   $+$0.10    &  $+$0.03  &  $+$0.06  &    0.00     &   $+$0.01  &     0.00     &   $+$0.01 &   $+$0.01  &   $+$0.04 \\ \hline
Total                      &   0.18    & 0.10         &  0.09     &  0.14     &    0.07      &  0.18     &    0.06      &
0.18    &    0.12    &   0.16   \\           \hline
  \end{tabular} 
\end{minipage}
\end{table*}
 
\subsection{Carbon-enhanced stars}\label{Sect:CEMP} 

Studies of the CEMP stars are important for better understanding the Galaxy chemical evolution. High carbon abundance makes possible to measure lines of C~I in the visible spectral range even for very metal-poor (VMP) CEMP stars. \citet[][hereafter, BBL10]{2010AA...513A..72B} and \citet[][hereafter, SCB13]{2013AA...552A.107S} used C~I 4932~\AA, 5052~\AA, and 5380~\AA\ and the molecular CH lines to derive the carbon abundances of the four turn-off stars in the $-3.3 \le$ [Fe/H] $\le -2.5$ metallicity range from the 1D and 3D calculations with applying the NLTE abundance corrections for lines of C~I. They found large abundance discrepancies between lines of C~I and CH in both 1D and 3D analysis, with (C~I $-$ CH) of $-0.35$~dex to $-0.79$~dex in 1D and $-0.23$~dex to $+0.42$~dex in 3D.

 We noticed that the NLTE carbon abundance correction, $\Delta_{\rm NLTE} = -0.45$~dex, applied by BBL10 and SCB13 is rather large and performed the NLTE calculations using our model atom of C~I and atmospheric parameters from BBL10 and SCB13. The obtained NLTE effects turned out much smaller, with $\Delta_{\rm NLTE} = -0.04$~dex. Using the 1D-LTE abundances from BBL10 and SCB13 and our $\Delta_{\rm NLTE}$s, we found consistent abundances from the two chemical species for three of four investigated stars, with (C~I $-$ CH) = 0.02~dex (BBL10's star) and $-0.03$~dex and 0.07~dex (SCB13's stars). For one of the BBL10's stars (C~I $-$ CH) = $-0.46$~dex. In 3D the difference (C~I $-$ CH) is positive everywhere, up to $+0.7$~dex.



This motivated us to compute the NLTE abundance corrections for lines of C~I in the small grid of model atmospheres appropriate for the CEMP stars. We adopted $S_{\rm H}$ = 0.3 and $\xi_t$ = 1.5~\kms. The results are presented in Table~\ref{tab00}. For the visible lines, C~I 4932\,\AA, 5052\,\AA, 5380\,\AA, and 6587\,\AA, $\Delta_{\rm NLTE}$ nowhere exceeds 0.13~dex in absolute value. The NLTE corrections are more negative, up to $-0.70$~dex, for the IR lines at 9061\,\AA, 9078\,\AA, and 9111\,\AA. The SE calculations were performed with [C/Fe] = 1, 2, and 3, although the employed MARCS model atmospheres \citep{2008AA...486..951G} were computed with [C/Fe] = 0. To check how $\Delta_{\rm NLTE}$ depends on the carbon abundance used in the model atmosphere construction, we asked Frank Grupp to calculate the two MAFAGS-OS \citep{Grupp2004b, Grupp2009} models with common atmospheric parameters, 6250/4.0/$-3$, but different [C/Fe] = 0 and [C/Fe] = 3. When employing a common carbon abundance of [C/Fe] = 3 in the SE calculations, we found very similar NLTE effects for C~I in these two models.

\begin{table}
  \caption{ NLTE abundance corrections for C~I lines in the carbon-enhanced model atmospheres. }
        \label{tab00} 
  \begin{tabular}{lccccccc}\hline \hline  
$\lambda$, \AA\ &  4932 & 5052  & 5380   &  6587  &  9061 &  9078 &  9111  \\                                                  \hline
log~$g$         & \multicolumn{7}{c}{ \Teff\ = 4500~K, [M/H] = $-$3, [C/Fe] = 2}     \\  \hline  
0.5             & -0.10 & -0.12 & -0.12  & +0.01  & -0.26 & -0.21 & -0.26    \\
1.0             & -0.08 & -0.09 & -0.10  & 0.00   & -0.21 & -0.17 & -0.21   \\
2.0             & -0.06 & -0.06 & -0.07  & -0.02  & -0.13 & -0.10 & -0.13   \\  \hline                                                                      
                & \multicolumn{7}{c}{ \Teff\ = 5000~K, [M/H] = $-$2, [C/Fe] = 2}     \\  \hline
1.5             & -0.08 & -0.12 & -0.10  &  -0.06 & -0.55 & -0.50 & -0.59  \\
2.0             & -0.06 & -0.09 & -0.07  &  -0.06 & -0.48 & -0.41 & -0.50  \\
2.5             & -0.04 & -0.07 & -0.05  &  -0.05 & -0.37 & -0.32 & -0.39  \\  \hline
                & \multicolumn{7}{c}{ \Teff\ = 5500~K, [M/H] = $-$2, [C/Fe] = 2}     \\  \hline
2.0             & -0.08 & -0.13 & -0.10  & -0.09  & -0.66 & -0.59 & -0.70  \\
3.0             & -0.05 & -0.06 & -0.04  & -0.05  & -0.44 & -0.40 & -0.49  \\
4.0             & -0.02 & -0.03 & -0.02  & -0.02  & -0.21 & -0.20 & -0.25  \\  \hline                                             
                & \multicolumn{7}{c}{ \Teff\ = 6250~K, [M/H] = $-$2, [C/Fe] = 1}     \\  \hline                                                                      
3.0             & -0.02 & -0.02 & -0.02  & -0.03  & -0.24 & -0.19 & -0.26  \\
4.0             & -0.01 & -0.01 & -0.01  & -0.01  & -0.14 & -0.10 & -0.15  \\
5.0             & -0.01 & -0.01 & -0.01  & -0.01  & -0.06 & -0.05 & -0.06  \\  \hline 
                & \multicolumn{7}{c}{ \Teff\ = 6250~K, [M/H] = $-$3, [C/Fe] = 3}     \\  \hline  
3.0             & -0.05 & -0.10 & -0.07  & -0.07  & -0.65 & -0.55 & -0.69   \\
3.5             & -0.04 & -0.06 & -0.05  & -0.05  & -0.54 & -0.45 & -0.56   \\
4.0             & -0.03 & -0.05 & -0.03  & -0.04  & -0.41 & -0.34 & -0.39   \\
4.5             & -0.02 & -0.03 & -0.02  & -0.02  & -0.29 & -0.24 & -0.29  \\
5.0             & -0.01 & -0.02 & -0.01  & -0.01  & -0.19 & -0.14 & -0.19  \\  \hline 
  \end{tabular}
\end{table}

\subsection{Comparison with previous studies}\label{Sect:Comparison}

\citet{1992AJ....104.1568T} determined carbon abundances of 34 metal-poor halo dwarfs 
from the atomic C~I and molecular CH lines and found that the atomic lines lead to systematically higher abundances compared with those from the CH lines, with the average difference, [C/Fe]$_{\rm CI} -$ [C/Fe]$_{\rm CH}$ = +0.4 dex. 
The C~I lines were analysed based on the NLTE line formation.
For four stars in common with that study 
we obtained [C/Fe]$_{\rm CI} -$ [C/Fe]$_{\rm CH} = -0.02$~dex from our calculations and +0.25~dex from the data of \citet{1992AJ....104.1568T}.
The discrepancy between the two studies can be due to using a 130~K, on average, lower effective temperature in \citet{1992AJ....104.1568T} compared with our work. For the common stars we employed the IRFM effective temperatures, while \citet{1992AJ....104.1568T} the temperatures based on the $b-y$, $R-I_J$, $R-I_K$, and $V-K$ colors. It can be seen from Table\,\ref{error} that using the lower temperature leads to the higher C~I-based and the lower CH-based abundance. 

\section{Conclusions}\label{Sect:conclusions}

We constructed a comprehensive model atom for C~I using the most up-to-date atomic data and 
evaluated the NLTE line formation for C~I in classical 1D models representing the atmospheres of late-type stars, where carbon abundance varies from solar value down to [C/H] = $-3$.

In line with the previous studies we found that NLTE leads to stronger C~I lines compared with their LTE strengths and negative NLTE abundance corrections.
The deviations from LTE are large for the strong lines, C~I 9061-9111~\AA\ (multiplet 3), 9405~\AA, and 9658~\AA, that form in the uppermost atmospheric layers, where collisions are inefficient. For these lines $\Delta_{\rm NLTE}$s range between $-0.10$~dex and $-0.45$~dex depending on stellar parameters. The NLTE corrections do not exceed 0.03~dex in absolute value for the weaker visible lines C~I 4932~\AA, 5052~\AA, 5380~\AA, and 6587~\AA. The NLTE effects depend strongly on the carbon abundance in the atmosphere. Therefore, the NLTE abundances of individual stars should be determined using an appropriate carbon abundance in the SE calculations.


In the $-1 \le$ [M/H] $\le 0$ range our theoretical results agree well with that of \citet{2005PASJ...57...65T} and \cite{2006AA...458..899F}. At [M/H] = $-2$ our $\Delta_{\rm NLTE}$s are consistent with that of \citet{2005PASJ...57...65T}, but they are 0.08~dex smaller in absolute value compared with that of \citet{2006AA...458..899F}. We computed smaller NLTE abundance corrections for the [M/H] = $-3$ model atmospheres than the corresponding values in \citet{2005PASJ...57...65T} and \cite{2006AA...458..899F}. This is not due to applying the \cite{Reid1994} data. We did not find any observational data in the literature to check, what is correct. 

The treated model atom was applied to analyse lines of C~I in the Sun and reference late-type stars 
covering the $-2.56 \le$ [Fe/H] $\le -1.02$ range. The solar and stellar carbon abundances were also determined from the molecular CH lines, using recent molecular data from \citet{2014AA...571A..47M}. We aimed to check, whether different abundance indicators lead to consistent results for each star.

The solar abundances derived from lines of C~I, CH, and C$_2$ are consistent within the error bars, when applying $S_{\rm H}$ = 0.3 to 1 in the SE calculations for C~I. Our results are in line with the 1D and 3D solar abundances of \citet{2009ARAA..47..481A} and \citet{2010AA...514A..92C}.

For each program star the difference between the NLTE ($S_{\rm H}$ = 0.3) abundance from the atomic C~I lines and the molecular CH lines does not exceed 0.07~dex. An exception is a cool VMP dwarf BD~$+66^\circ$268, where the atomic IR lines are rather weak resulting in (C~I - CH) = 0.17~dex. The obtained results suggest that the molecular CH lines can safely be used for stellar carbon abundance determinations with classical 1D model atmospheres in a broad metallicity range, at least, down to [Fe/H] $\simeq -2.4$. 

We show that the NLTE abundance corrections for C~I 4932~\AA, 5052~\AA, and 5380~\AA\ in the carbon-enhanced 6200/4.0/$-3$ models do not exceed 0.04~dex in absolute value. Applying our $\Delta_{\rm NLTE}$ to the carbon 1D-LTE abundances of the four CEMP stars as determined by \citet{2010AA...513A..72B} and \citet{2013AA...552A.107S} leads to (C~I $-$ CH) of smaller than 0.07~dex for three of them. In 3D the difference (C~I $-$ CH) is positive everywhere, up to $+0.7$~dex. For the 4th star neither 1D nor 3D helps to make abundances from the different chemical species consistent. We present the NLTE abundance corrections for lines of C~I in the carbon-enhanced models of different \Teff, log~$g$, and [M/H]. They will be useful for accurate determinations of the carbon abundances of the CEMP stars.

The NLTE method treated in this study for C~I will be used in our further studies for accurate determinations of not only stellar carbon abundances, but also effective temperatures. The abundance difference between lines of C~I and CH can serve as an efficient \Teff\ indicator. 


{\it Acknowledgements.}
We thank Frank Grupp for calculating the MAFAGS-OS carbon-enhanced model atmospheres and Tatyana Ryabchikova for helpful discussion.
This work was supported by the Russian Foundation for Basic Research (grants 14-02-31780 and 14-02-91153).




\bibliography{carbon}
\bibliographystyle{mn2e}

\end{document}